\documentclass[12pt,a4paper,aps,showpacs,showkeys,amssymb,pra,floatfix,preprint,nofootinbib]{revtex4}

\usepackage[fleqn,tbtags,intlimits]{amsmath}

\usepackage{slashed}
\usepackage{xcolor}
\usepackage{hyperref}
\usepackage{booktabs}
\usepackage[squaren]{SIunits}

\usepackage[pdftex]{graphicx}
\usepackage{grffile}
\usepackage{feynmp}

\DeclareSymbolFont{rsfs}{U}{rsfs}{m}{n}
\DeclareSymbolFontAlphabet{\mathrsfs}{rsfs}
\usepackage[mathscr]{eucal}
\usepackage{bbold}

\newcommand{\phice}{\phi_{\rm CE}}
\renewcommand{\d}{\mathrm d}

\renewcommand{\mathbf}{\boldsymbol}

\newcommand{\enflux}{ \Phi }

\begin{document}

\title{Asymmetries of azimuthal photon distributions in non-linear Compton scattering in
ultra-short intense laser pulses}
\author{D.~Seipt}
\email{d.seipt@hzdr.de}
\altaffiliation[Present address: ]{Helmholtz-Institute Jena, Friedrich Schiller Universit\"at Jena, 07743 Jena, Germany}

\author{B.~K{\"a}mpfer}
\affiliation{Helmholtz-Zentrum Dresden-Rossendorf, P.O.~Box 510119, 01314 Dresden, Germany}
\affiliation{TU Dresden, Institut f{\"u}r Theoretische Physik, 01062 Dresden, Germany}


\date{\today}

\begin{abstract}

Non-linear Compton scattering
in ultra-short intense
laser pulses is discussed with
the focus on angular distributions of the emitted photon energy.
This is an observable
which is accessible easily experimentally. Asymmetries of
the azimuthal distributions are predicted for both linear and
circular polarization.
We present a systematic survey of the influence
of the laser intensity, the carrier envelope phase and the laser polarization on
the emission spectra for single-cycle and few-cycle laser pulses.
For linear polarization, the dominant direction of the emission changes from
a perpendicular pattern with respect to the laser polarization at low-intensity 
to a dominantly parallel emission for high-intensity laser pulses.

\end{abstract}

\maketitle

\section{Introduction}

The development of novel bright short-pulsed X-ray radiation sources is an important issue with respect to applications in materials research, dynamics investigations and biological structure ranging up to medical application \cite{Carpinelli:NIMA2008}.
At the heart of these table-top X-ray sources is the inverse
Compton process, where optical laser photons are scattered off relativistic electrons,
possibly laser accelerated ones \cite{Debus:APB2010},
and Doppler up-shifted to the X-ray regime.
While quasi-monochromatic X-rays may be achieved using long laser pulses at
rather low intensities,
pulsed broadband X-ray sources can be developed with high-intensity short pulse lasers,
on the contrary.
Besides the application-oriented research, the laser-particle interaction offers a
rich variety of interesting perspectives also for fundamental physics of
particle dynamics and radiation processes in strong electromagnetic fields, both experimentally and theoretically \cite{DiPiazza:RevModPhys2012}.
For instance, the formation of QED avalanches at
ultra-high laser intensities
is related to the emission of high-energy photons
off ultra-relativistic electrons and subsequent pair production induced by those very
photons \cite{Bell:PRL2008,Fedotov:PRL2010}.

Here, we focus on the non-linear Compton scattering process, where photons from an intense laser pulse ($L$) scatter off a relativistic free electron ($e$), emitting a single non-laser photon $\gamma'$ 
in the reaction $e + L \to e' +\gamma' + L$.
Multi-photon interactions due to the large photon density in the laser pulse give rise
to non-linear effects such as the emission of high harmonics 
and the intensity-dependent red-shift of the scattered radiation~\cite{DiPiazza:RevModPhys2012}.
Each of the harmonics shows a characteristic multi-pole pattern,
which in fact was used for an experimental identification of the higher harmonics
\cite{Chen:Nature1998,Babzien:PRL2006}.
Non-linear strong-field effects for
Compton scattering, and also for the related process of pair production, have been observed in the SLAC E-144 experiment \cite{Bula:PRL1996,Bamber:PRD1999}.
The detailed theoretical and experimental knowledge of the radiation spectra
might allow to use the emitted photon radiation as a diagnostics tool
to measure, e.g.,~the laser intensity \cite{DiPiazza:OptLett2012},
or to determine the carrier envelope phase of the laser pulse \cite{Mackenroth:PRL2010},
or the parameters of the electron beam \cite{Leemans:PRL1996}.

The theoretical investigations of non-linear Compton scattering started
shortly after the invention of the laser in a series of seminal papers
\cite{Nikishov:JETP1963,Nikishov:JETP1964a,Nikishov:JETP1964b,
Narozhnyi:JETP1964,Goldman:PhysLett1964,Brown:PR1964,Kibble:PR1965}
(see also \cite{Harvey:PRA2009} for a complete overview of the literature
as well as the reviews \cite{Ehlotzky:ProgPhys2009,DiPiazza:RevModPhys2012}).
While these early papers paved the way for further investigations, they mostly
specified monochromatic infinite plane waves to model the laser field since at
that time long picosecond and nanosecond laser pulses were common.
However, nowadays the use of ultra-short, femtosecond laser pulses has become
standard due to the development of chirped pulse amplification which
allows to reach new high-intensity frontiers with lasers such as
ELI \cite{site:ELI}
envisaging intensities as high as $\unit{10^{24}}{\watt\per\centi\metre^2}$ in pulses of
$\unit{15}{\femto\second}$.
Even shorter pulse lengths will become available with the
petawatt field synthesizer (PFS) \cite{Major:AIPCP2010}, producing high intensities with pulse lengths of the order of $\unit{5}{\femto\second}$, which corresponds to less than two laser cycles.

In such short laser pulses, the Compton spectra are drastically altered
\cite{Narozhnyi:JETP83,Boca:PRA2009,Seipt:PRA2011,Mackenroth:PRA2011,Boca:EPJD2011}
as compared to the previously common treatment by means of infinitely long plane waves.
Important effects are the ponderomotive broadening of the harmonics
and the appearance of substructures,
anisotropies in the angular spectra as well as the relevance of the carrier envelope phase \cite{Lan:JPB2007,Krajewska:PRA2012}.
For instance,
it has been shown that the value of the carrier envelope phase determines the
angular region of emission \cite{Mackenroth:PRL2010}.
Due to the large bandwidth of the emitted radiation it is in
general not possible to distinguish individual harmonics in strong pulsed fields.
Therefore, in this paper, we discuss energy-integrated angular spectra in
ultra-short intense laser pulses. We give a systematic survey
of the influence of the laser pulse parameters on the angular spectra,
focussing on asymmetries in azimuthal distributions of the emitted energy.

Our paper is organised as follows:
After a concise presentation of the Volkov states and the basic framework,
we derive the analytical expressions for the differential emission probability and energy distribution
in Section \ref{sect:amplitude}.
In Section \ref{sect:numerics}, we present a comprehensive numerical study of the
azimuthal spectra in ultra-short laser pulses and the dependence of the photon energy distributions
on the laser pulse parameters.
In Section \ref{sect:energy} we discuss the total amount of emitted energy in relation to the
primary energy flux in the laser pulse.
The conclusions are drawn in Section \ref{sect.conclusion}.
In Appendix \ref{app:class} we briefly discuss the classical framework, that is Thomson scattering, for the angular photon spectra.
We find that the total emitted energy is proportional to
the integrated primary energy flux in the laser pulse.
In Appendix \ref{app:pert}, we derive the non-relativistic and ultra-relativistic limits the azimuthal cross sections for Compton scattering of a polarized photon in perturbative QED.
It is shown that, in the ultra-relativistic case, the emission is azimuthally symmetric in the leading order.

\section{Scattering amplitude, probability and emitted energy}
\label{sect:amplitude}

To account for the strong laser pulse non-perturbatively, one can work within the
Furry picture and employ Volkov wavefunctions as a basis for the perturbative
expansion of the $S$ matrix. The Volkov states \cite{Volkov:1935} are solutions of the Dirac equation in the presence of the plane-wave background field\footnote{Natural units with $\hbar = c = 1$ are employed throughout this paper,
as well as the Feynman dagger notation $\slashed p = \gamma_\mu p^\mu$.} $A_\mu$
\begin{align}
(i\slashed\partial - e \slashed A(\phi) - m )\Psi(x) = 0 \,.
\label{eq:dirac}
\end{align}
We consider here a laser pulse described by the transverse
($A\cdot k \equiv A_\mu k^\mu=0$) vector potential
\begin{align}
A^\mu(\phi) = A_0 g(\phi) {\rm Re\,} \big( \epsilon^\mu_+ e^{-i(\phi + \phice)} \big) \,,
\label{eq:vector.potential}
\end{align}
depending on the invariant phase $\phi = k\cdot x$ with the laser wave four-vector $k^\mu=(\omega,0,0,-\omega)$.
The polarization of the background field is described by complex polarization vectors $\epsilon^\mu_\pm = \delta^\mu_1 \cos \xi \pm i \delta^\mu_2 \sin \xi$
(the quantity $\delta^\mu_\nu$ denotes the Kronecker symbol),
with polarization parameter $\xi$.
Here, $\xi=0,\pi/2$ means linear polarization and $\xi= \pi/4$ denotes circular polarization; all other values refer to an arbitrary elliptic polarization.
In \eqref{eq:vector.potential}, $g(\phi)$ denotes the pulse envelope.
The limit $g\to1$ refers to infinitely long plane wave fields (IPW), while a finite support 
of $g(\phi)$ describes a pulsed plane wave (PPW) to be specified below.
The relative phase between the pulse envelope and the carrier wave is
the carrier envelope phase (CEP) $\phice$.
The dimensionless laser amplitude $a_0$, which quantifies relativistic and multi-photon effects, is defined as $a_0 = |e| A_0/m$, where $A_0$ denotes the amplitude of the
vector potential \eqref{eq:vector.potential},
and $e = -|e| = - \sqrt{4\pi \alpha}$ is the charge of the electron
with the fine structure constant $\alpha \simeq 1/137$.
The values of $a_0$ are related to the peak laser intensity $I$ and the laser central wavelength $\lambda$ via $a_0^2 = \unit{7.3\times 10^{-19}}
{ I[\watt\per\centi\metre^{2}] \lambda^2[\micro\metre]} $.

The solutions of Eq.~\eqref{eq:dirac} are given by \cite{Ritus:JSLR1985}
\begin{align}
\Psi_{pr}(x) &=
\left(
1 + \frac{e}{2k\cdot p} \slashed k \slashed A
\right)
\exp
\left\{ -i p\cdot x
-\frac{i}{2k\cdot p} \int_0^\phi \d \phi' ( 2 e p \cdot A - e^2 A\cdot A )
\right\} u_p \,,
\label{eq:ritus.matrix}
\end{align}
with the free spinor $u_{pr}$, normalized to $\bar u_{pr} u_{pr'} = 2m \delta_{rr'}$.

\begin{figure}[!t]
\begin{center}
\begin{fmffile}{feynman_nlcompton}
$${
	\parbox{40mm}{
	\begin{fmfgraph*}(40,17)
	 \fmfleft{l,d}
	 \fmfright{r,d}
	 \fmftop{d,d,t1,t2,d}
	 \fmf{dbl_plain,tension=1.25}{l,v1}
	 \fmf{dbl_plain,label.side=right}{v1,r}
	 \fmffreeze
	 \fmf{photon}{v1,t2}
	 \fmfdot{v1}
	 \fmflabel{$k',\lambda'$}{t2}
	 \fmflabel{$p,r$}{l}
	 \fmflabel{$p',r'$}{r}
	 \end{fmfgraph*}}
	 }
$$
\end{fmffile}
\end{center}
\caption[First-order Feynman diagram for non-linear one-photon Compton scattering.]{
First-order Feynman diagram for non-linear one-photon Compton scattering within the Furry picture. In strong-field QED, the one-photon Compton scattering appears
as the decay of a laser dressed Volkov electron state (straight double lines) with momentum $p$
and spin quantum number $r$ into another laser dressed electron with momentum $p'$ and spin $r'$ while emitting a photon (wavy line) with momentum $k'$ in a polarization state $\lambda'$.
}
\label{fig:feynman.diagram} 
\end{figure}
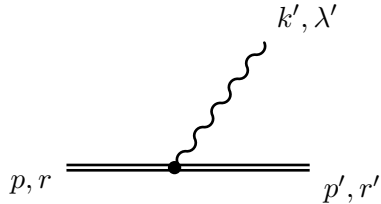

The Volkov states Eq.~\eqref{eq:ritus.matrix}, which include the interaction
of the laser pulse with the electrons, are used as in- and out-states for the electrons when
calculating matrix elements. The interaction with non-laser photons is treated in perturbation theory,
depicted in the Feynman diagram in Fig.~\ref{fig:feynman.diagram}
for non-linear Compton scattering in a strong laser field, i.e.~the emission of a non-laser mode photon off a Volkov electron. Thus, non-linear Compton scattering
is a first-order process ``above'' the non-perturbative interaction of the electron with the laser field.
The $S$ matrix for the process is obtained by using the corresponding Feynman rules \cite{Mitter:1975} as
\begin{align}
S & 
= -ie \int \! \d^4x \, \bar \Psi_{p'r'}(x) \slashed \epsilon'_{\lambda'} e^{ik'\cdot x} \Psi_{pr}(x) \,.
\end{align}
Performing the spatial integrations over the light-front variables\footnote{The light-front components of a four-vector $x^\mu$ are defined as
$x^\pm = x^0 \pm x^3$ and $\mathbf x^\perp = (x^1,x^2)$
with the scalar product $x\cdot y = (x^+y^- + x^-y^+)/2 - \mathbf x^\perp \cdot \mathbf y^\perp$. The  four-dimensional volume element reads $\d^4x = \d x^+ \d x^- \d^2 \mathbf x_\perp / 2$.
Note that $k^-=2\omega$ is the only non-vanishing light-front component of the laser four-vector $k^\mu$ and, therefore, $\phi = \omega x^+$.}
$\mathbf x^\perp$ and $x^-$
one arrives at
\begin{align}
S & = -ie (2\pi)^4 \int \! \frac{\d s}{2\pi} \, \delta^{(4)}(p' + k' - p- sk) \mathrsfs M(s)\,,
			\label{eq:matrix.element.2} \\
\mathrsfs M(s) &= T_0 \mathrsfs C_0(s) 
				+ T_+ \mathrsfs C_+(s)
				+ T_- \mathrsfs C_-(s)
				+ T_2 \mathrsfs C_2(s)
\end{align}
with Dirac current structures
\begin{align}
T_0 &= \bar u_{p'r'} \slashed \epsilon' u_{pr} \,,\\
T_\pm &
		=  \bar u_{p'r'}\left( d_{p'} \slashed \epsilon_\pm \slashed k \slashed \epsilon'
		+  d_p \slashed \epsilon' \slashed k \slashed \epsilon_\pm \right) u_{pr}
		\,,\\
T_2 &
        = d_p d_{p'}  (\epsilon'\cdot k) \bar u_{p'r'} \slashed k u_{pr} \,,
\end{align}
and $d_{p^{(\prime)}} = ma_0/(2k\cdot p^{(\prime)})$. The variable $s$ in \eqref{eq:matrix.element.2} parametrizes the momentum transfer between the background laser field and the electron by means of the momentum
conservation $p + sk = p'+k'$.
It might be interpreted as net number of laser photons absorbed by the electron only in the IPW case due to the periodicity of the background field which allows
to interpret the discrete Fourier components of the classical background field as ``quanta'' with momentum $k^\mu$.
Thus, in the IPW case the energy resolved photon spectrum consists of discrete harmonics, while
in the PPW case a smooth continuous spectrum emerges.

Evaluating the integral over $s$ in \eqref{eq:matrix.element.2} fixes
the value $s \equiv (k'\cdot p)/(k\cdot p')$ as a function of the frequency and polar angle of the emitted photon.
This equation may be inverted to yield
\begin{align}
\omega'(s,\vartheta) &=
\frac{s k\cdot p}{(p+sk)\cdot n'}
= \frac{s \omega e^{2\zeta}}{ 1 + e^{\zeta} \sinh \zeta (1-\cos\vartheta)+ s \frac{\omega e^\zeta}{m} (1+\cos\vartheta)}
\end{align}
with $n' = (1,\mathbf n')$, the unit vector
$\mathbf n' = (\sin \vartheta \cos \varphi, \sin\vartheta,\sin\varphi,\cos\vartheta)$,
i.e.~$k'=\omega'n'$, and the initial electron rapidity $\zeta = {\rm Arcosh}\, \gamma$.
The energy-momentum conservation is reduced to the conservation of three light-front
components of momentum, $p^+ = p'^+ + k'^+$ and $\mathbf p_\perp = \mathbf p'_\perp + \mathbf k'_\perp$.

The integrals over the laser phase determining the amplitude of the process are given by
\begin{align}
\left\{
\begin{matrix}
\mathrsfs C_0(s) \\
\mathrsfs C_\pm(s) \\
\mathrsfs C_2(s)
\end{matrix}
\right\} &=
\int_{-\infty}^\infty \d \phi 
\, e^{is\phi - i f(\phi)}
\left\{
\begin{matrix}
1 \\
g(\phi) e^{\mp i(\phi+\phice)} \\
g^2(\phi) \Big( 1 + \cos 2\xi  \cos 2(\phi+\phice) \Big)
\end{matrix}
\right\}  \label{eq:Cn}
\end{align}
with
\begin{align}
f(\phi) &= \int_0^\phi \d \phi' \left\{ g(\phi') {\rm Re\,} \left[ \alpha_+  e^{-i(\phi' +\phice)} \right]
 + \beta  g(\phi')^2 \left[ 1 + \cos 2\xi \cos 2(\phi'+\phice)\right] \right\} \label{eq:f}
 \end{align}
and the coefficients
$\alpha_+ = d_p ( 2 \epsilon_+ \cdot p) -  d_{p'} ( 2 \epsilon_+ \cdot p')$ and
$ \beta = d_p^2 (k\cdot p) - d_{p'}^2 (k\cdot p') $.
The integral $\mathrsfs C_0(s)$
needs special a treatment as it is an integral over an infinite interval for a pure phase factor due to the
lacking pre-exponential pulse envelope function $g(\phi)$. It contains contributions
from the free electron motion outside the laser pulse leading to a divergent contribution at zero momentum transfer $s=0$.
For $s>0$, $\mathrsfs C_0$  can be related to the well defined integrals $\mathrsfs C_\pm$ and $\mathrsfs C_2$ --- which are rendered finite by the appearance of the pre-exponential pulse shape functions --- by the principle of gauge invariance \cite{Ilderton:PRL2011,Seipt:PRD2012}
\begin{align}
 s\mathrsfs C_0(s) = 
 \frac{\alpha_+}{2} \mathrsfs C_+(s)
+ \frac{\alpha_-}{2} \mathrsfs C_-(s)
+ \beta \mathrsfs C_2(s)
 \,.
\end{align}

The differential emission probability, depending on momentum as well as spin and polarization variables of the out-states, reads
$\d W_{rr'\lambda'} = |S|^2 \d \Pi / (2p^+)$  
with the Lorentz invariant phase space element
\begin{align}
\d\Pi = \frac{\d^3k'}{(2\pi)^3 2\omega'} \frac{\d^2{\mathbf p'}_\perp \d p'^+}{(2\pi)^3 2p'^+} \,,
\end{align} 
where the final electron phase space is conveniently parametrized in terms of
light-front variables due to the light-front momentum conservation.
Integrating over the final-electron variables with the momentum conservation in Eq.~\eqref{eq:matrix.element.2} one gets
\begin{align}
\d W_{rr'\lambda'} &= \frac{\alpha \omega' |\mathrsfs M(s ( \omega' ) )|^2}{16{ \pi^2} k\cdot p \, k\cdot p'} \, \d\omega' \, \d\Omega\,,
\end{align}
where $\d \Omega = \d\varphi \, \d \cos \vartheta$ is the solid angle element
related to the emitted photon of energy $\omega'$.
For numerical calculations, it is more convenient to employ $s$ as the independent variable, instead
of $\omega'$,
since then the three phase space integrals are independent from each other.
The transformation of the differentials reads
$\d\omega' = \frac{\omega'}{s} \frac{k\cdot p'}{k\cdot p} \d s$
and the phase space of the emitted photon is characterized by
$\varphi \in [-\pi,\pi)$, 
$\vartheta \in [0,\pi]$ and 
$s \in (0,\infty)$.

We are not interested in the spin and polarization states, so we average over the initial spin $r$
and sum over the final state spin and polarization variables, thus obtaining the differential probability
\begin{align}
w(s,\vartheta,\varphi) &\equiv \frac{1}{2} \sum_{r,r',\lambda'} \frac{\d W_{rr'\lambda'}}{\d s\d \Omega } \\
& = \frac{\alpha m^2 \omega'^2}{8\pi^2 (k\cdot p)^2 s}
\left\{
-2 |\mathrsfs C_0|^2  + \frac{a_0^2}{2} \left( 1 + \frac{u^2}{2(1+u)}\right) \right.  \nonumber \\
& \qquad
\left.
\vphantom{\frac11}  \times
\Big[
	|\mathrsfs C_+|^2 + |\mathrsfs C_-|^2 
	+ \cos 2 \xi \, \big( \mathrsfs C_+ \mathrsfs C_-^* + \mathrsfs C_- \mathrsfs C_+^* \big)
	- \mathrsfs C_0 \mathrsfs C_2^* - \mathrsfs C_2 \mathrsfs C_0^*
\Big]
\right\}   \label{eq:probability.2}
\end{align}
with the invariant variable $u =  (k\cdot k')/(k\cdot p')$.
In this expression, the azimuthal angle $\varphi$ appears solely via the
non-linear exponentials $f(\phi)$ of the functions $\mathrsfs C_n$ defined in \eqref{eq:Cn}.

The energy $E$ of the emitted radiation is given by the time component
of the momentum four-vector $P^\mu$ of the radiation field
\begin{align}
P^\mu &= \int \! \d s \, \d \Omega \, k'^\mu w(s,\vartheta,\varphi)\,. 
\end{align}
Note that the emitted energy $E$ and also the corresponding azimuthal distribution
$\d E/\d \varphi$ are not Lorentz invariant.
Instead, $E$ transforms under Lorentz boosts as the time component of the four-vector $P^\mu$,
$\bar{ E} = \hat \gamma ( E + \boldsymbol {\hat v} \cdot \mathbf {P})$,
where $\mathbf {\hat v}$ is the relative velocity and $\hat \gamma= (1-\mathbf {\hat v}^2)^{-1/2}$ is the corresponding Lorentz factor.
On the contrary, the total probability
\begin{align}
W &= \int \! \d s \, \d \Omega \, w(s,\vartheta,\varphi)
\label{eq:total.prob}
\end{align}
is a Lorentz invariant quantity.

\section{Numerical Analysis}

\label{sect:numerics}

We focus on parameters which
are accessible in various laboratories. To be specific, one parameter
set which may allow for an experimental verification
refers to an operating set-up
at the Helmholtz-Zentrum Dresden-Rossendorf (HZDR) using the electron linear accelerator ELBE \cite{Gabriel:NIMB2000} in combination with the short-pulse high-intensity laser DRACO \cite{Zeil:NJP2010}.
For the following numerical survey we specify head-on collisions
of an optical laser pulse ($\omega=\unit{1.55}{\electronvolt}$) with an electron beam which has an energy corresponding to $\gamma = 100$.

For weak laser fields, $a_0 \ll1$, the emission is dominantly into a narrow cone with a typical opening angle $\vartheta_{\rm cone} \sim 1/\gamma$,
centred at the initial velocity $\mathbf u_0$.
For larger laser strength $a_0>1$, the radiation cone widens due to the intensity dependent radiation pressure with a typical angle $\vartheta_{\rm cone} \sim a_0/\gamma$.
In particular, for a laser intensity of $a_0 = 2\gamma$ the mean emission angle is $\vartheta_{\rm cone} = \pi/2$ \cite{Harvey:PRA2009}.

In the following, we use the pulse envelope function
\begin{align}
g(\phi) & = 
\left\{
\begin{array}{ll}
\cos^{2} \left( \frac{\phi}{2N} \right)  & \text{for } \phi \in [-\pi N, \pi N]  \,,\\
0 															& \text{elsewhere} \,,
\end{array}
\right. 
\label{eq:envelope}
\end{align}
introduced in
Eq.~\eqref{eq:vector.potential} and entering the coefficient functions $\mathrsfs C_n$, with compact support on the interval $[-\pi N, \pi N]$. The number of optical cycles of the laser pulse is $N$.

\subsection{Double differential angular distribution}

We start our analysis by discussing the double differential energy distribution
\begin{align}
\frac{\d E}{\d \Omega} &= \int_0^\infty \d s \, \omega'(s,\vartheta) \, w(s, \vartheta, \varphi) \,.
\label{eq:double.diff}
\end{align}
We have in mind a high-granular pixel photon detector
placed in the radiation cone that allows to measure angularly resolved energy-integrated
spectra.
The quantity \eqref{eq:double.diff}, as energy-integrated emitted energy,
does not suffer from the possible problem of detector pile-ups,
where two individual photons are detected as a single photon of summed energy.
Such false detections may happen if the incident photon flux is too large for the
temporal resolution of the detector.
It provides, thus, an easily observable quantity to characterize appropriately the Compton spectrum.
The angular resolution can be improved by increasing the distance of the detector
from the interaction point of the electron beam and the laser pulse.

\begin{figure}
\includegraphics[width=0.32\textwidth]{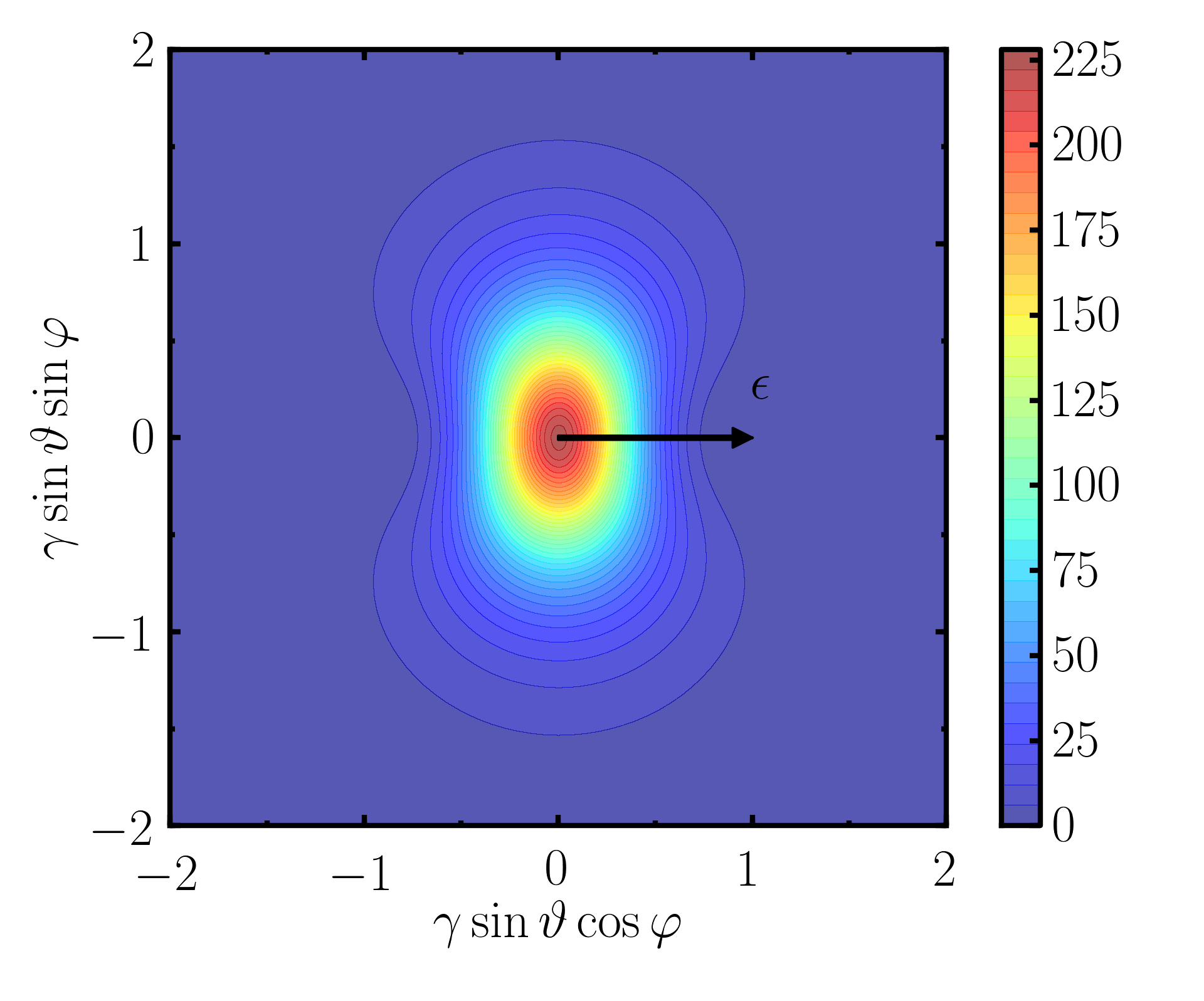}
\includegraphics[width=0.32\textwidth]{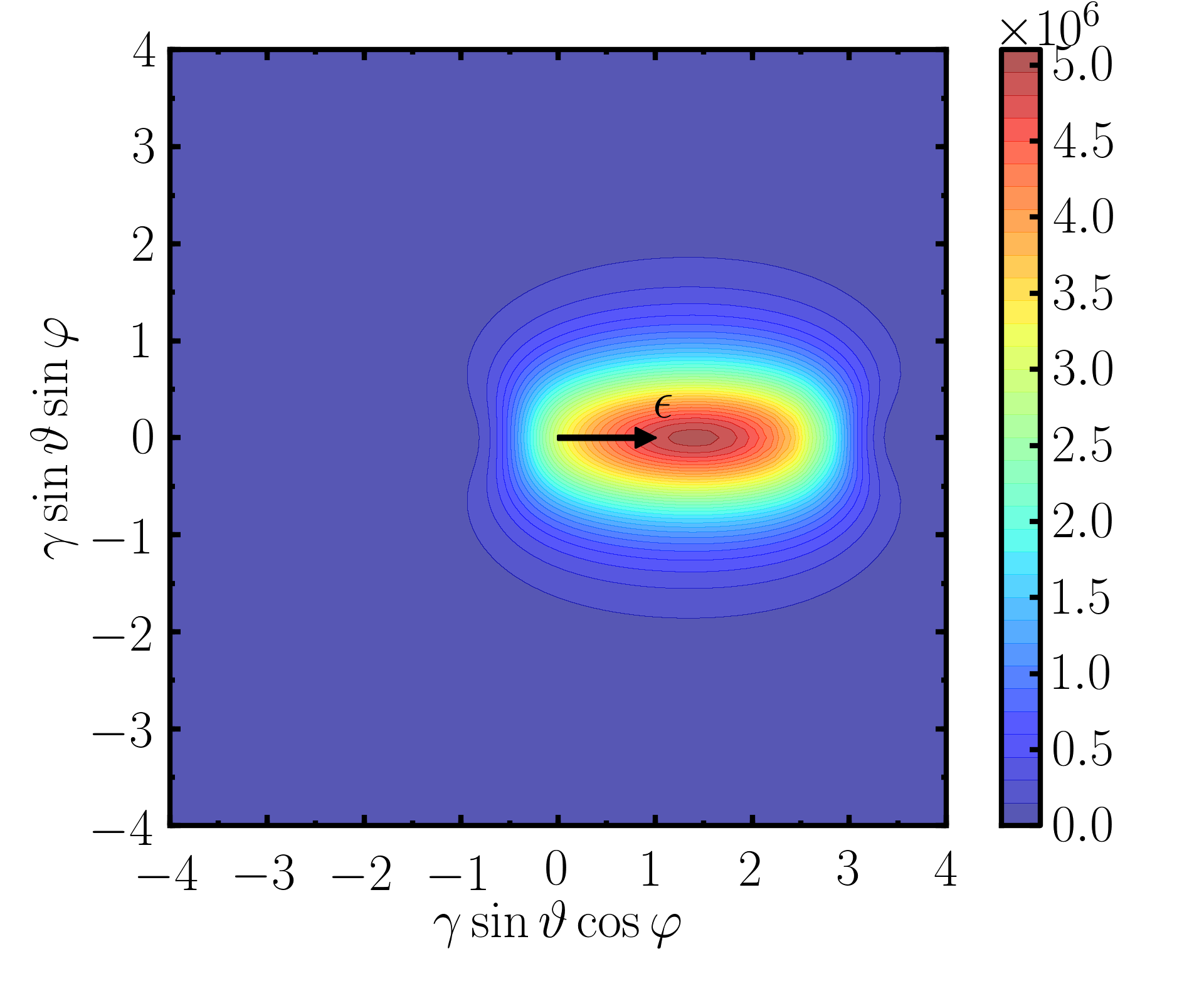}
\includegraphics[width=0.32\textwidth]{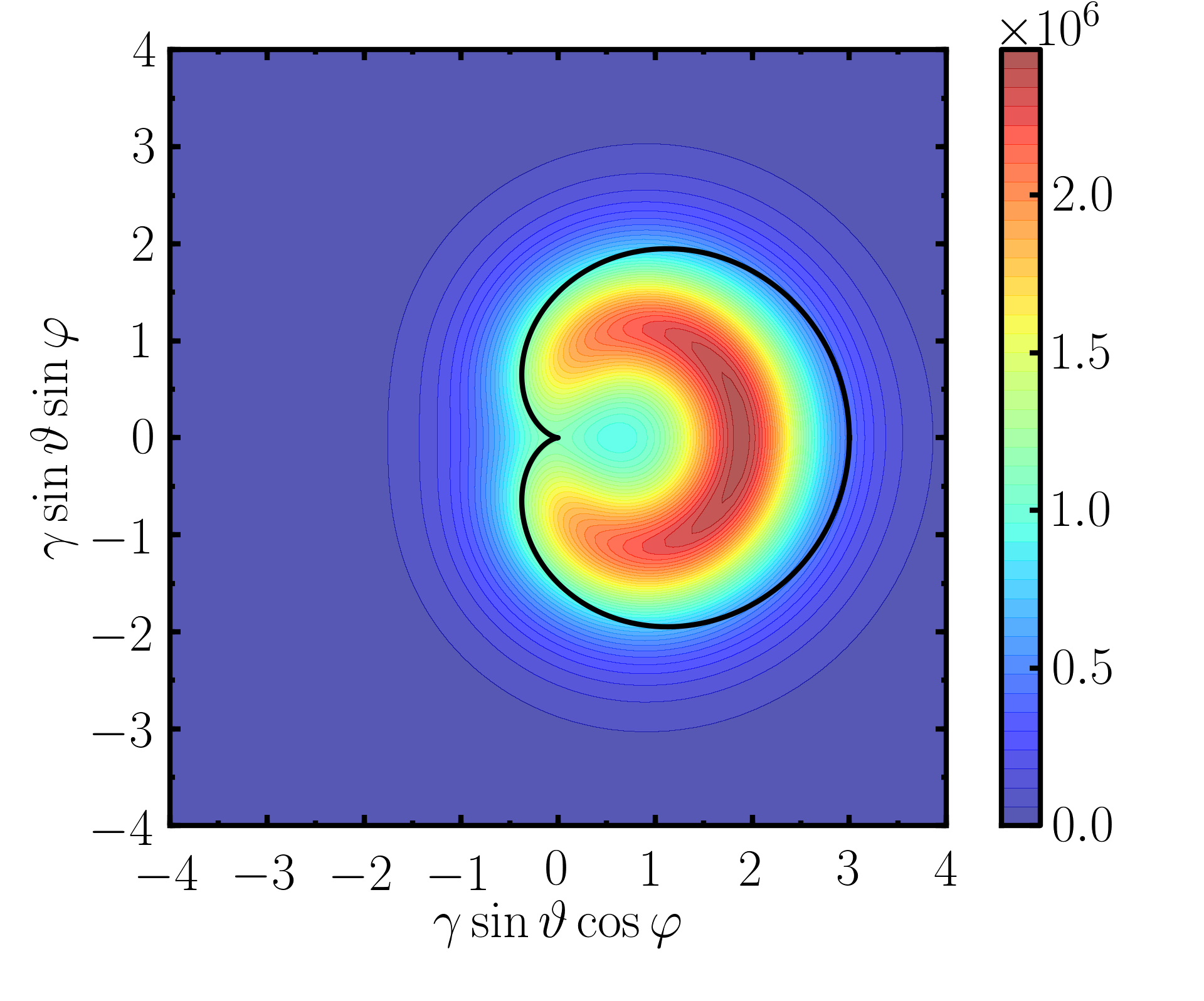}
\caption{Double differential angular photon energy distribution $\d E / \d \Omega$ for a
single-cycle ($N=1$) linearly polarized laser pulse with $\phice=0$ for $a_0=0.01$ (left panel) and $a_0=3$ (center panel).
The arrows denote the direction of the linear laser polarization vector $\boldsymbol \epsilon$.
In the right panel, the result for circular polarization is shown for $a_0=3$. In that case, the black curve depicts the amplitude of the laser vector potential in the $x-y$ plane.}
\label{fig:double.differential}
\end{figure}

Figure~\ref{fig:double.differential} exhibits a series of the distributions $\d E/ \d \Omega$ as
contour plots. 
At low laser intensity, $a_0 \ll1$, the shape of the distribution
is oriented perpendicularly to the polarization vector, which is $\mathbf e_x$ in our case.
This is the usual dipole-type emission.
For increasing values of $a_0>1$, the shape of the spectrum flips and is oriented
in the direction of the polarization vector.
This transition is characteristic for the non-linear interactions. The details of the
spectra for field strengths $a_0>1$, i.e.~whether it is
an unidirectional emission as in the central panel of Fig.~\ref{fig:double.differential}
or a bi-polar emission in the direction of the laser polarization,
depends strongly on further pulse parameters.
For ultra-short pulses this is mainly dominated by the value of the CEP.
These details are studied in the next subsection using the azimuthal distributions.

In the right panel of Fig.~\ref{fig:double.differential}, the corresponding
angular distribution is exhibited for a circularly polarized laser pulse with $a_0=3$.
Also in this case, strong asymmetries arise. The distribution peaks in
the direction where the laser vector potential (depicted as black curve) reaches its maximum value
and almost vanishes in the opposite direction.
The assertion of the radiation having an opening angle $\sim a_0/\gamma$ for $a_0>1$ is clearly verified by the angular distributions in Fig.~\ref{fig:double.differential}.

\subsection{Azimuthal distributions}

The information on the angular distribution of the emitted energy can be condensed
into the azimuthal distribution by integrating over the polar angle via
\begin{align}
\frac{\d E}{\d\varphi} &
 = \intop_0^\infty  \!\d s \intop_0^\pi \! \d\vartheta \,
 \sin \vartheta \, \omega'(s,\vartheta)\, w(s,\vartheta,\varphi)  \,.
\end{align}

For long pulses and IPW laser fields, the azimuthal spectra show a characteristic multi-pole pattern
for each harmonic, which have been observed by choosing a single harmonic
using appropriate energy filters \cite{Chen:Nature1998,Babzien:PRL2006}.
In the case of a short intense laser pulse, the distinction of different harmonics is not possible.
We thus study, as in the previous subsection, the energy-integrated spectrum.
In the following, the azimuthal distribution $\d E / \d \varphi$ of the emitted energy
will be analysed systematically.

\subsubsection{Dependence on the laser intensity}

\begin{figure}[!t]
\begin{center}
\includegraphics[width=0.24\textwidth]{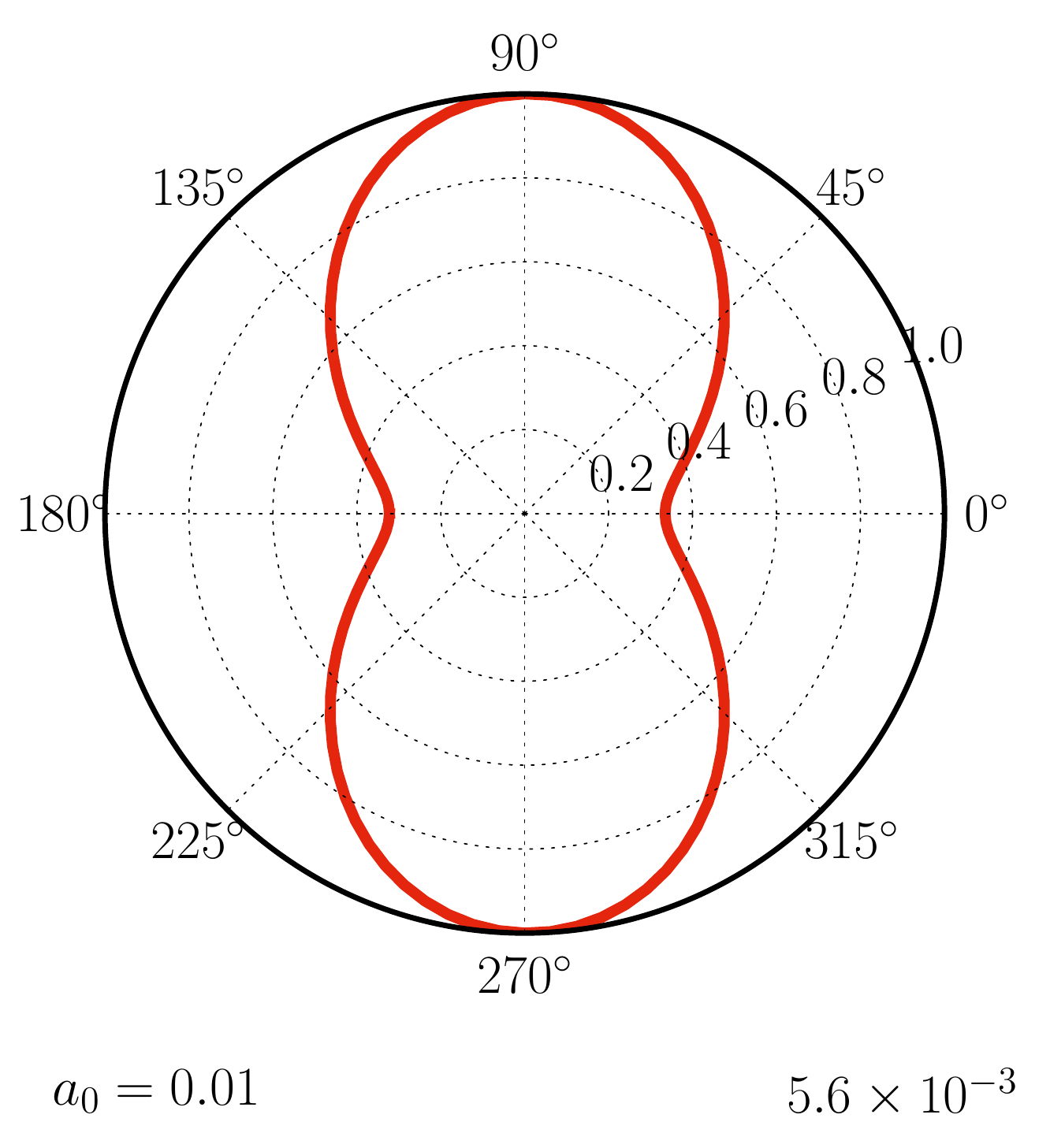}
\includegraphics[width=0.24\textwidth]{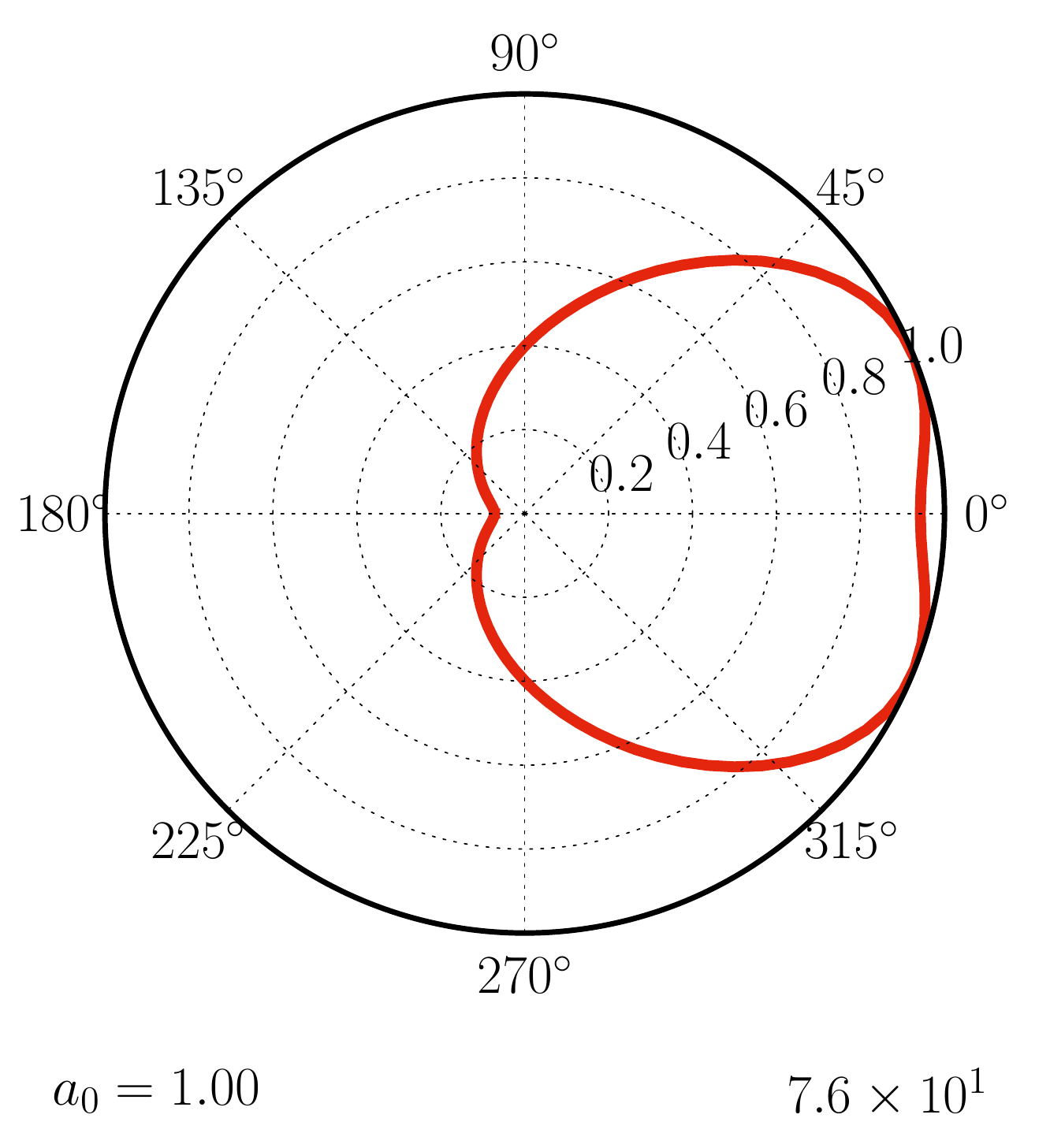}
\includegraphics[width=0.24\textwidth]{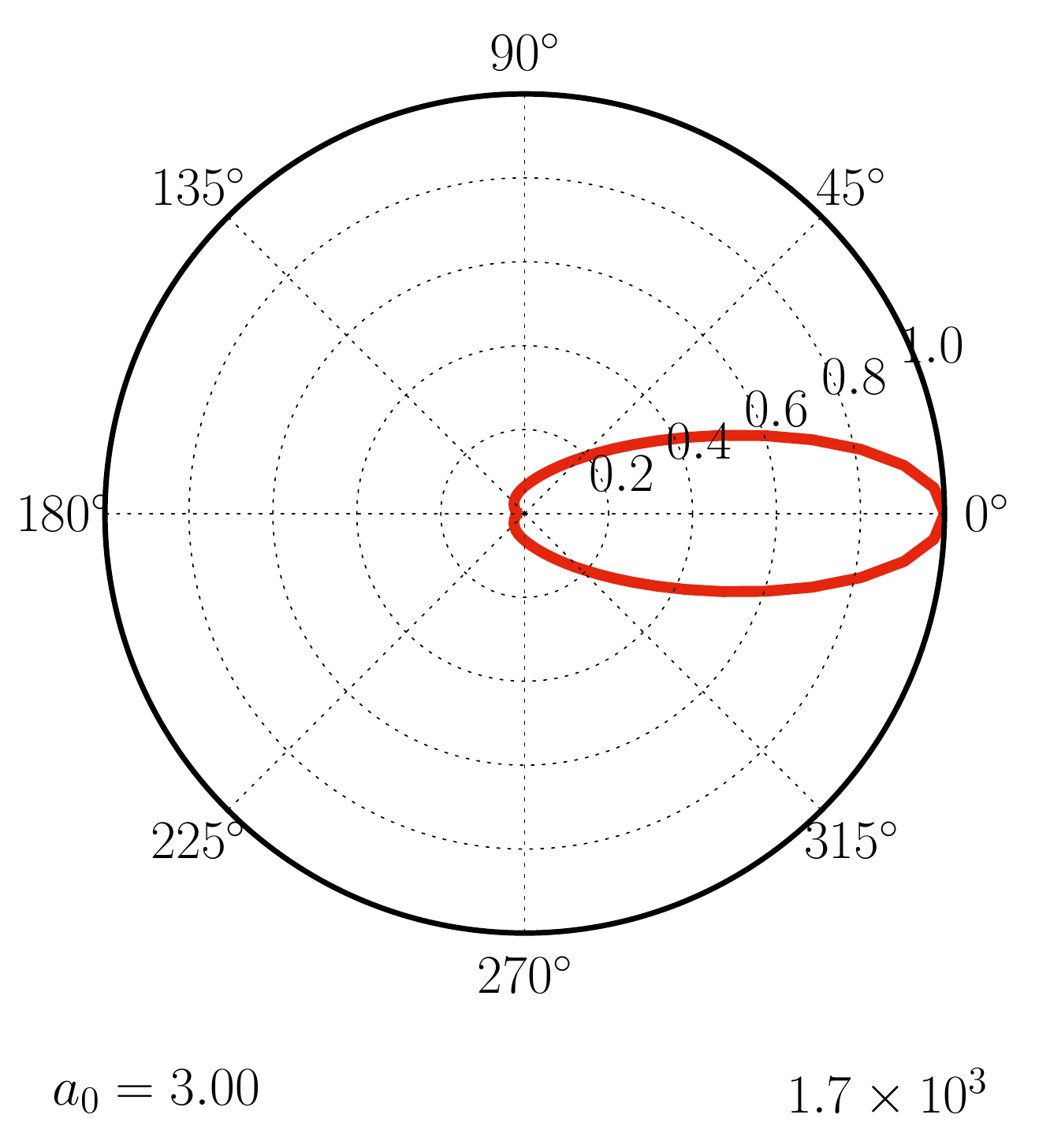}
\raisebox{0.45cm}{\includegraphics[width=0.24\textwidth]{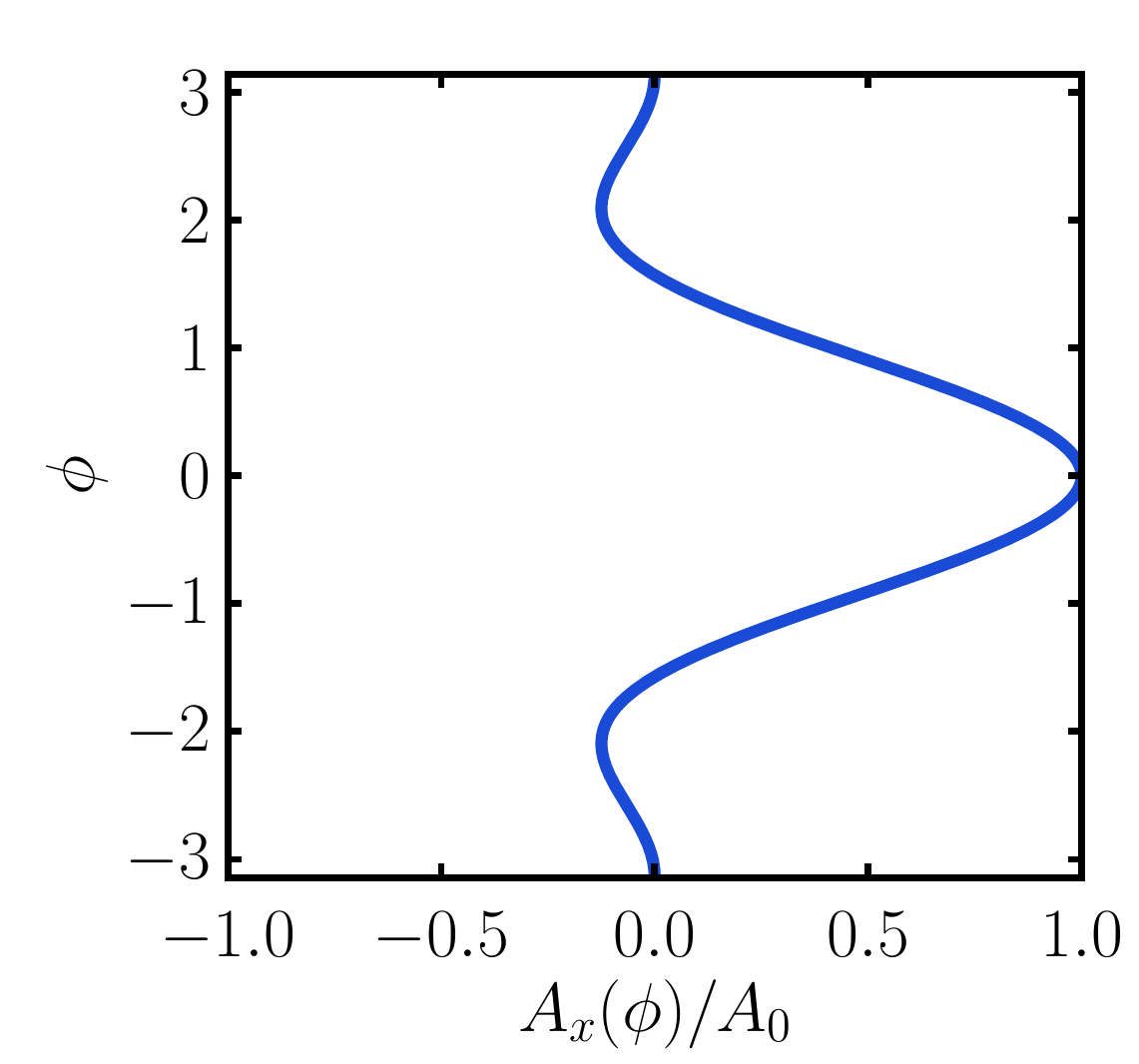}}
\end{center}
\caption{The normalized azimuthal distributions of emitted energy $\d E/\d\varphi$ for non-linear Compton scattering as a function of the azimuthal angle $\varphi$
in a single-cycle laser pulse with $N=1$ and $\phice=0$ for linear polarization ($\xi=0$)
for various values of $a_0 = 0.01$, $1$ and $3$ (from left to right).
In the rightmost panel, the laser vector potential $A^\mu$ is plotted as a function of the laser phase $\phi$ (note the inverted axes).
The labels at each panel give the maximum values of $\d E/\d\varphi$ in units of
$\electronvolt$.
The calculations have been performed in the laboratory frame where
the angle $\vartheta$ was integrated over a cone with opening angle $3/\gamma$
around the initial electron momentum.}
\label{fig:survey.linear}
\end{figure}

\begin{figure}
\includegraphics[width=0.49\textwidth]{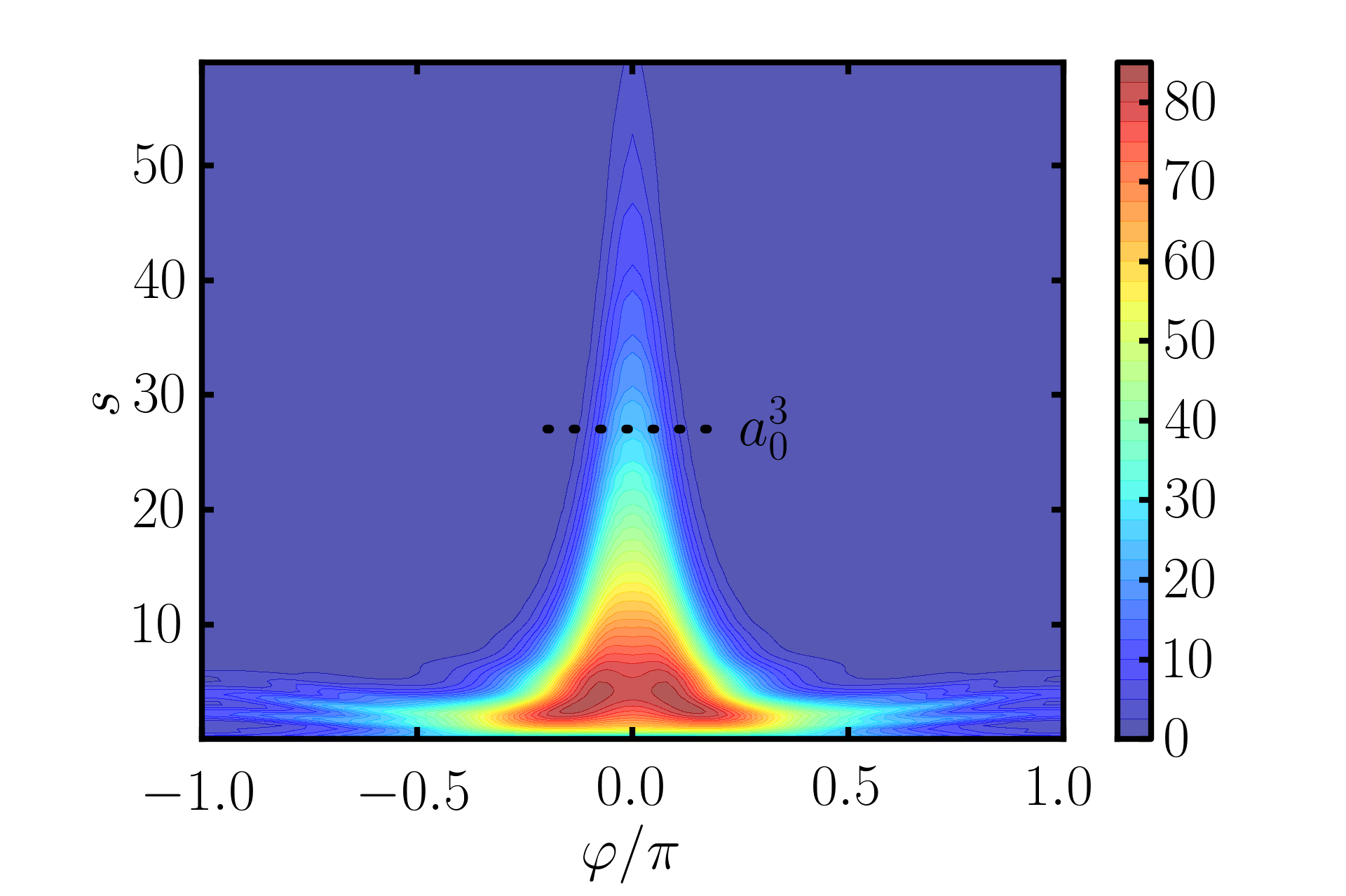}
\caption{Contour plot of the
energy resolved spectrum $\d E/ \d\varphi \d s$ over the $\varphi$-$s$ plane. The calculation is for $N=1$, $a_0=3$ and $\phice=0$. The typical range of values of $s$ is of the order of $a_0^3$.}
\label{fig:s.theta}
\end{figure}

\begin{figure}[!t]
\begin{center}
\includegraphics[width=0.24\textwidth]{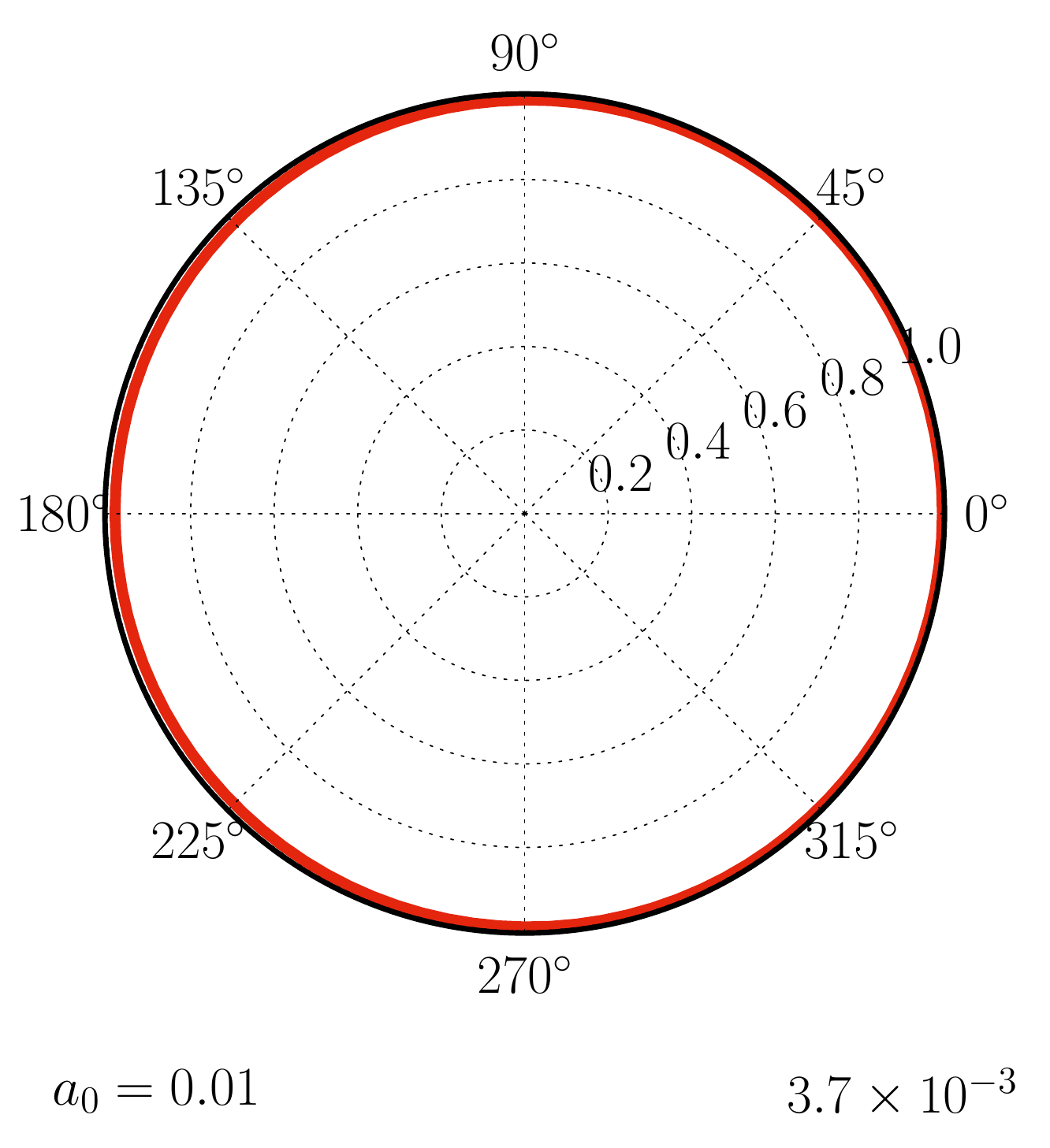}
\includegraphics[width=0.24\textwidth]{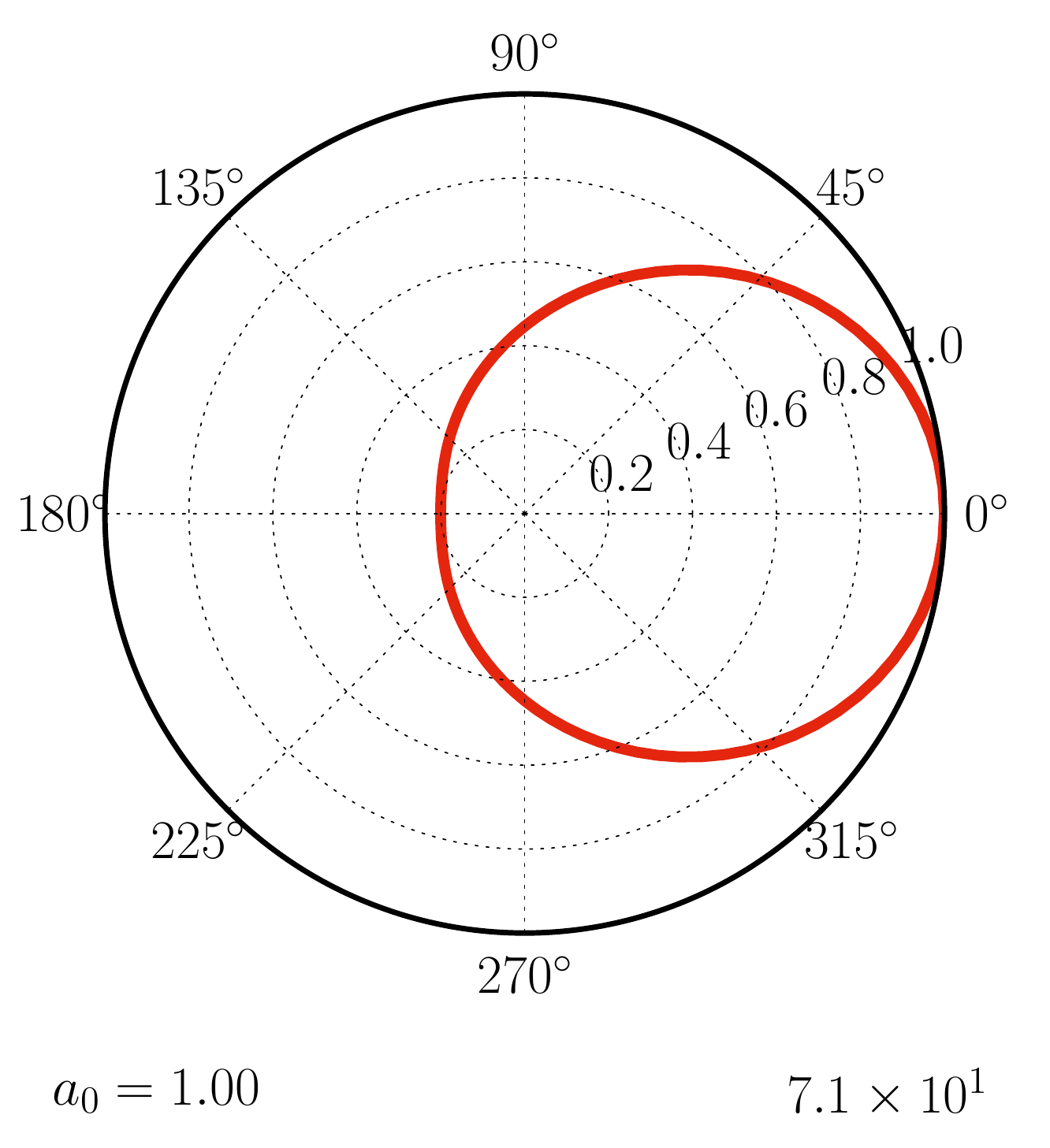}
\includegraphics[width=0.24\textwidth]{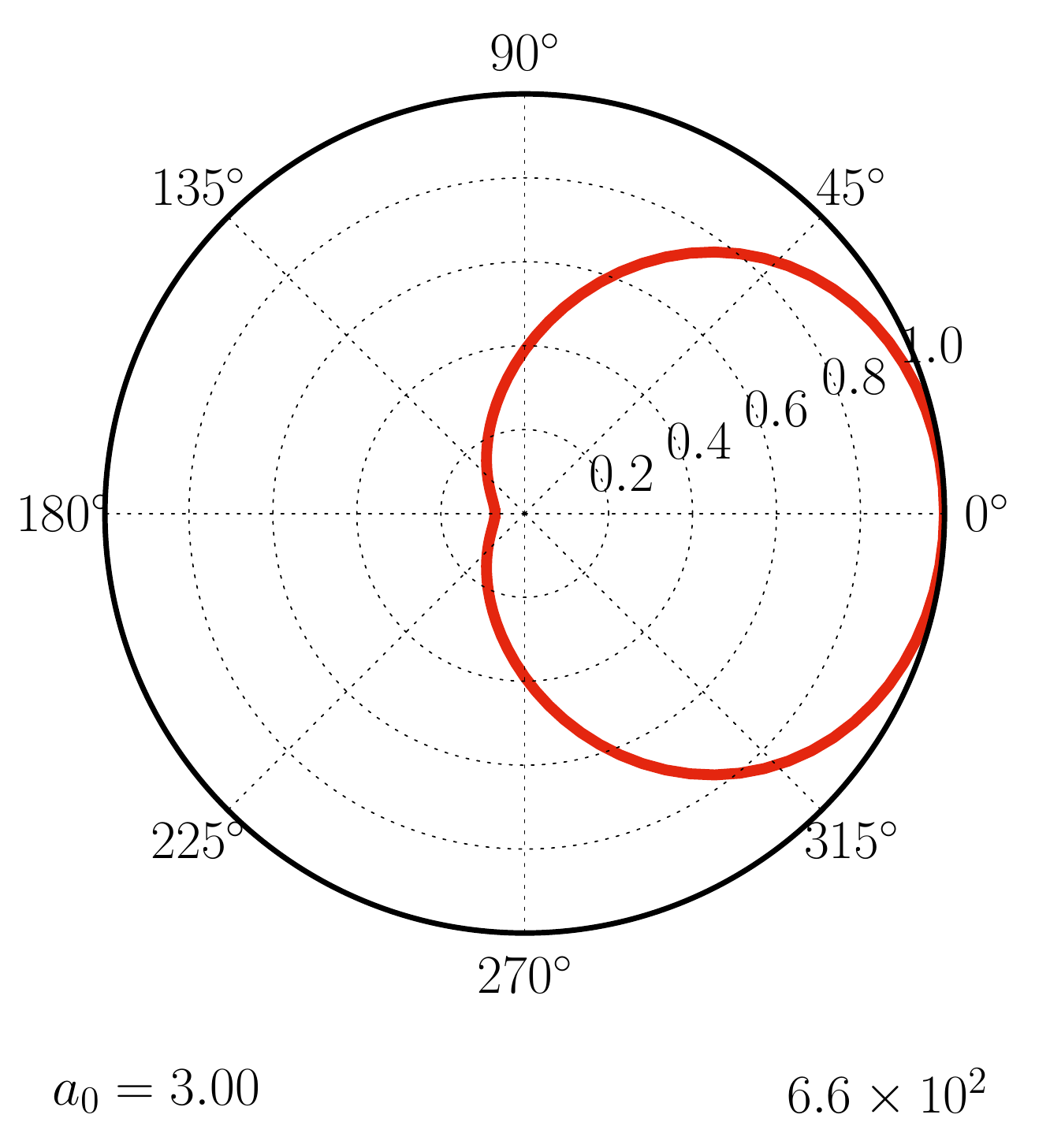}
\raisebox{0.45cm}{\includegraphics[width=0.24\textwidth]{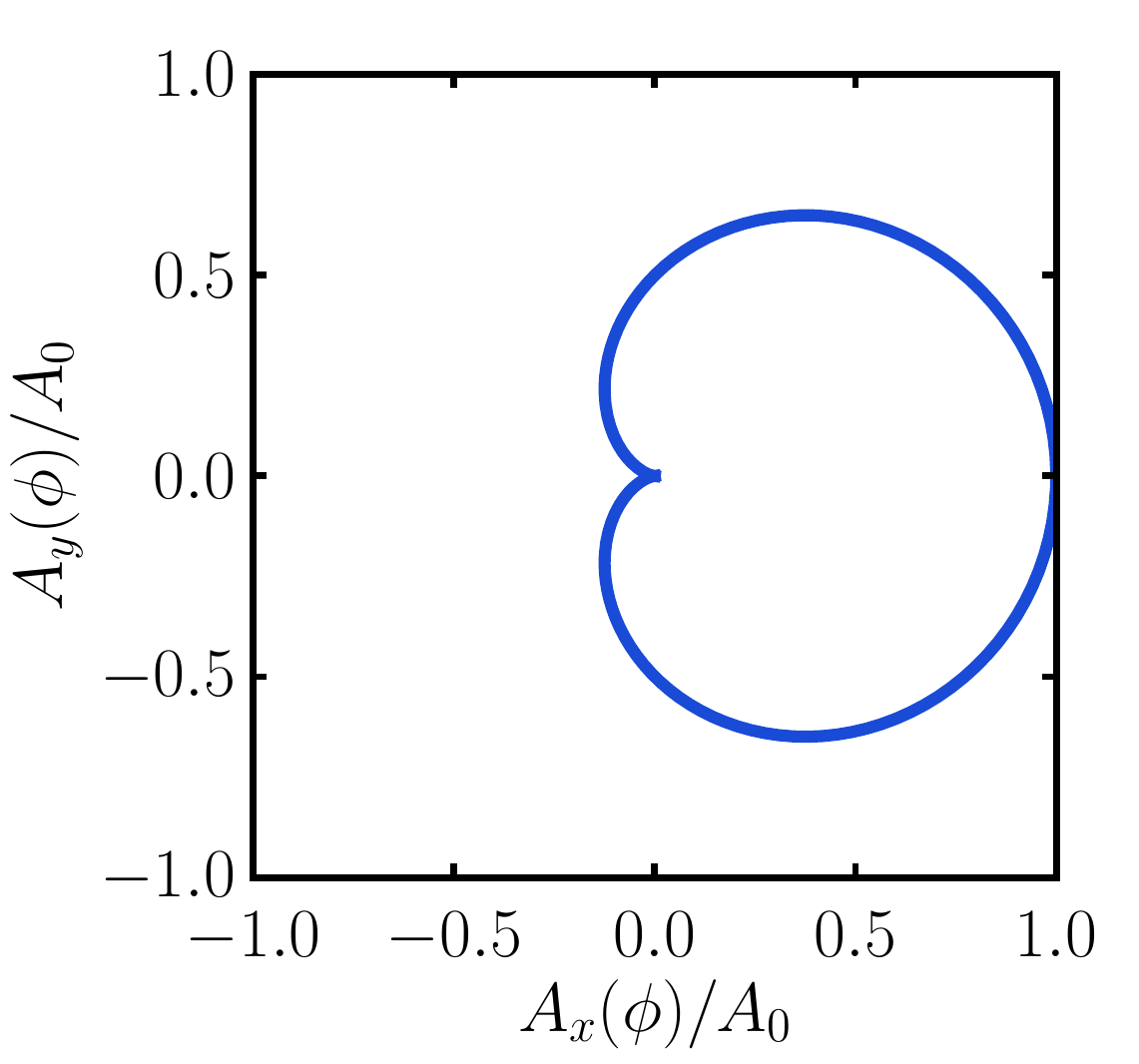}}
\end{center}
\caption{Azimuthal energy emission spectra as in Fig.~\ref{fig:survey.linear} but for circular
laser polarization with $\xi= \pi/4$. In the rightmost panel, the values of the laser vector potential
are plotted in the $x$-$y$ plane. The direction $\varphi=0$ is distinguished by the maximum
of the vector potential.}
\label{fig:survey.circular}
\end{figure}

The differential azimuthal energy spectra
$\d E/\d\varphi$ are exhibited in Fig.~\ref{fig:survey.linear} for a pulse length $N=1$, i.e.~for a single-cycle laser pulse.
At low laser intensity, $a_0=0.01$ (left panel), the emission has a strong dipole pattern with 
the preferred emission in the plane transverse to the polarization of the laser, which is here chosen as the $x$ axis, i.e.~$\varphi \sim 0^\circ,180^\circ$.
Upon increasing the laser strength up to $a_0=1$
and $a_0=3$ (middle and right polar diagrams), the shape of the spectrum develops towards an unidirectional
emission in the direction of the laser polarization.
The azimuthal spectra show the same qualitative behaviour as the double differential angular distributions discussed
above.
This is
explained by the behaviour of the vector potential, depicted in the rightmost panel of Fig.~\ref{fig:survey.linear}, which has a strong asymmetry and acquires large values only
in the direction $\varphi=0$. The typical range of values for the variable
$s$ scales as $a_0^3$ \cite{Ritus:JSLR1985}, as depicted in Fig.~\ref{fig:s.theta}.
However, due to the large asymmetry of the vector potential, the value of $a_0$, which refers to the peak value of $A^\mu$, is relevant only in the direction where the maximum occurs, which is
for $\varphi=0$ in our case.
That means in
the direction $\varphi=0$ high-energetic photons with large values of the
variable $s$ are emitted,
while in the opposite direction, $\varphi=\pi$, only low-energy photons are emitted, see Fig.~\ref{fig:s.theta}.
Since we are considering here the emitted energy,
the emission of high-energy photons with large values of the variable $s$ is weighted stronger than
it would be the case for the emission probability. Thus, the azimuthal
distributions of the emitted energy $\d E/\d \varphi$ are more sensitive to
asymmetries of the vector potential than the corresponding
azimuthal emission probabilities $\d W/\d \varphi$.

In Fig.~\ref{fig:survey.circular}, the azimuthal spectrum $\d E /\d\varphi$
is exhibited in polar plots for a circularly polarized single-cycle ($N=1$) laser pulse with $\xi=\pi/4$ and increasing values of $a_0$ from left to right, showing the asymmetry in strong laser pulses.
This is due to the fact that the vector potential has no azimuthal symmetry
for an ultra-short pulse, since a distinguished direction is defined by the maximum
of the laser pulse vector potential. As in the case of linear polarization, only in that
direction where $A^\mu$ reaches its maximum value, high frequency photons with
large values of the variable $s$ are emitted.
The behaviour of the vector potential in the azimuthal plane is depicted in the rightmost
panel of Fig.~\ref{fig:survey.circular}.
	
A weak short laser pulse with $a_0 \ll1$ does not show strong asymmetry effects.
This asymmetry is a combined
short-pulse intensity effect mainly due to the directional emission of high harmonics
with $s \gg 1$ (see Fig.~\ref{fig:s.theta}).
The transition from a distribution which is peaked perpendicularly to
the laser polarization (see Fig.~\ref{fig:survey.linear}, left polar diagram) to a distribution peaked in the direction of the laser polarization
vector (see Fig.~\ref{fig:survey.linear}, middle polar diagram) is characteristic for increasing the value of $a_0$ below unity to larger than unity.
The details of the shape of the distribution for $a_0>1$, i.e.~whether one observes a dipole-type
pattern or an uni-directional emission, depends on further pulse shape parameters.
These dependencies are studied in the following.

Note that the photon distribution
is determined by the shape and the symmetries of the laser vector potential, and not
by the electric field (see also the discussion in \cite{Krajewska:PRA2012c}), since
the Volkov wave functions \eqref{eq:ritus.matrix}
as well as the classical electron velocity $u^\mu$ in Eq.~\eqref{eq:orbit} both depend
directly on the laser vector potential $A^\mu$.
 
\subsubsection{Dependence on the pulse length}

\begin{figure}[!t]
\begin{center}

\includegraphics[width=0.49\textwidth]{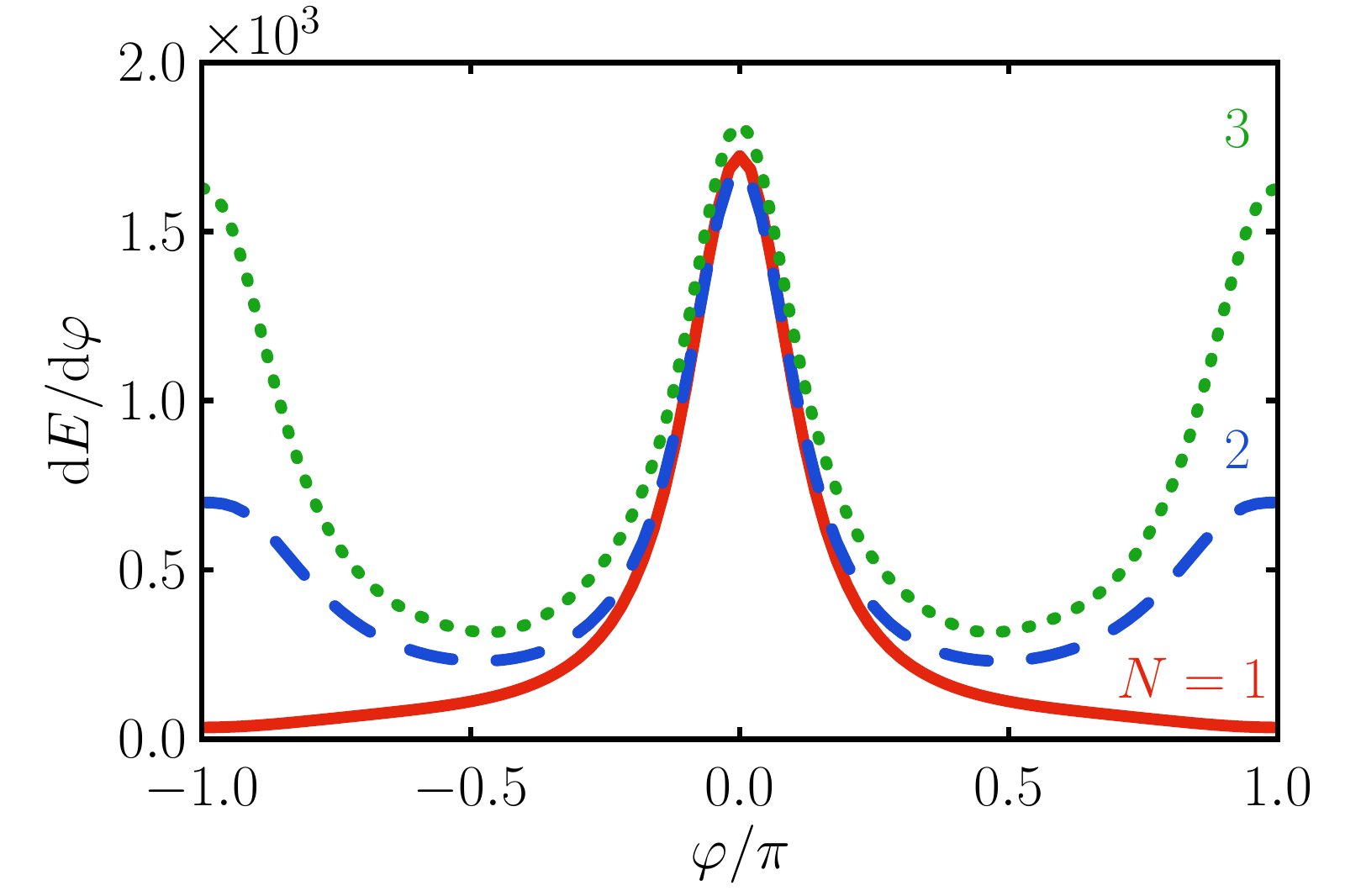}
\llap{\raisebox{0.195\textwidth}{\includegraphics[width=0.160\textwidth]{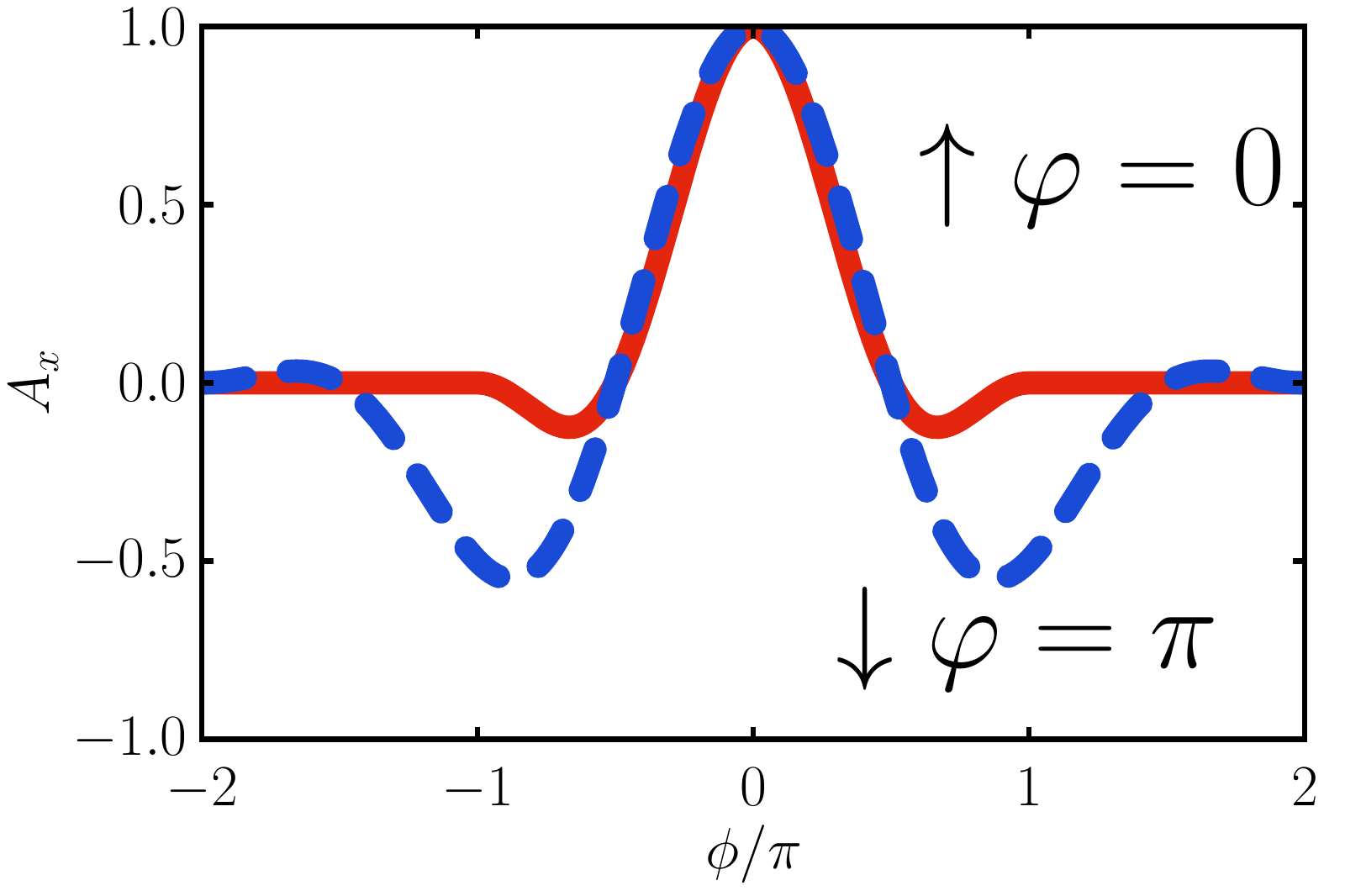}}
\hspace{0.245\textwidth} } 
\includegraphics[width=0.49\textwidth]{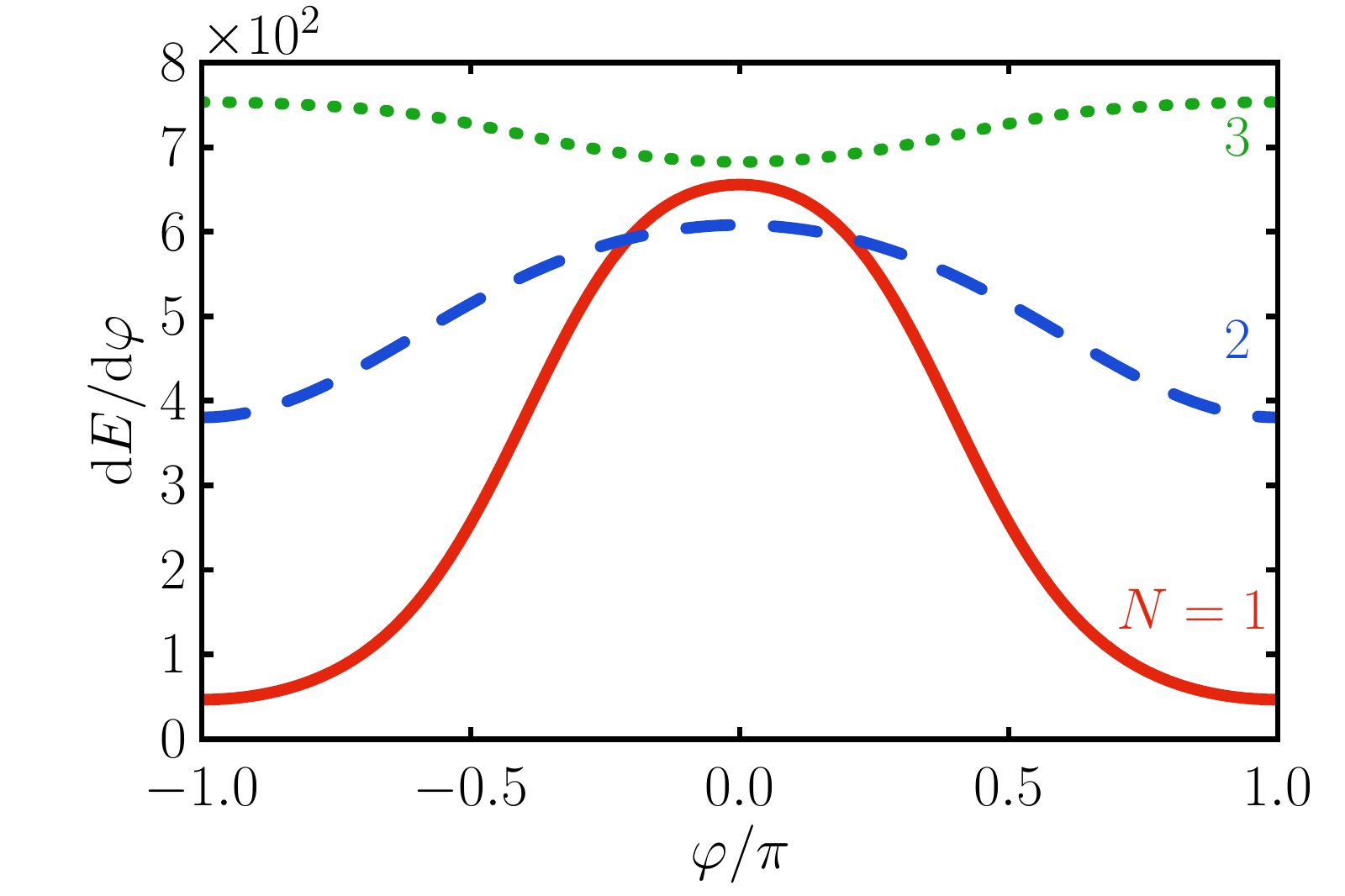}

\end{center}
\caption{Pulse length dependence of the azimuthal distributions $\d E / \d \varphi$ for non-linear Compton scattering
for linear (circular) polarization in the left (right) panel for $a_0=3$ and $\phice=0$. The line styles represent
	$N=1$ (red solid), 
	$2$ (blue dashed), 
	$3$ (green dotted).
}
\label{fig:pulselength}
\end{figure}

The above discussion was for single-cycle laser pulses with various strengths. Here, we now present
a systematic survey of the pulse length dependence of the azimuthal emission spectra for fixed values of $a_0$.
For low laser intensity, $a_0 \ll1$, the shape of the azimuthal distribution is independent of
the pulse length $N$.
On the contrary, in strong laser pulses the shape of the spectra strongly depends on
the pulse length.
While for $N=1$ the emission pattern in a strong laser pulse is highly asymmetric for a linearly polarized laser pulse, it is expected to become $\pi$-periodic for long pulses, with the limit
being the result of an infinite plane wave, where the
electron's quiver motion also becomes periodic.
Increasing the pulse length from $N=1$ to $2$ oscillations
adds to the azimuthal spectrum only in the direction $\varphi=\pi$, since the vector potential
at the center of the pulse, responsible for the emission in the direction $\varphi=0$,
remains almost unchanged, see left panel of Fig.~\ref{fig:pulselength}.

Going further to $N=3$, the results for the azimuthal spectra show a considerable restoration of that left-right symmetry. However, there is still a measurable difference of the order of $12\%$. For even longer pulses, the symmetry restoration proceeds further, such that for pulses longer than $N=5$ the differences are reduced to the level of below $1 \%$ and slowly decreases further
for even longer pulses. Thus, for long pulses one observes a bi-polar emission
pattern in the direction of the laser polarization.

Analogously, 
the strong azimuthal anisotropies
must disappear for longer laser pulses in the case of circular laser polarization.
For still rather short pulses with $N=3$ the azimuthal isotropy is recovered partly,
see right panel of Fig.~\ref{fig:pulselength}. Interestingly,
the maximum of the spectrum for $N=1$ develops into the minimum for $N=3$.
For very long pulses one recovers azimuthal isotropy.

\subsubsection{Dependence on the carrier envelope phase}

\begin{figure}[!t]
\begin{center}
\includegraphics[width=0.49\textwidth]{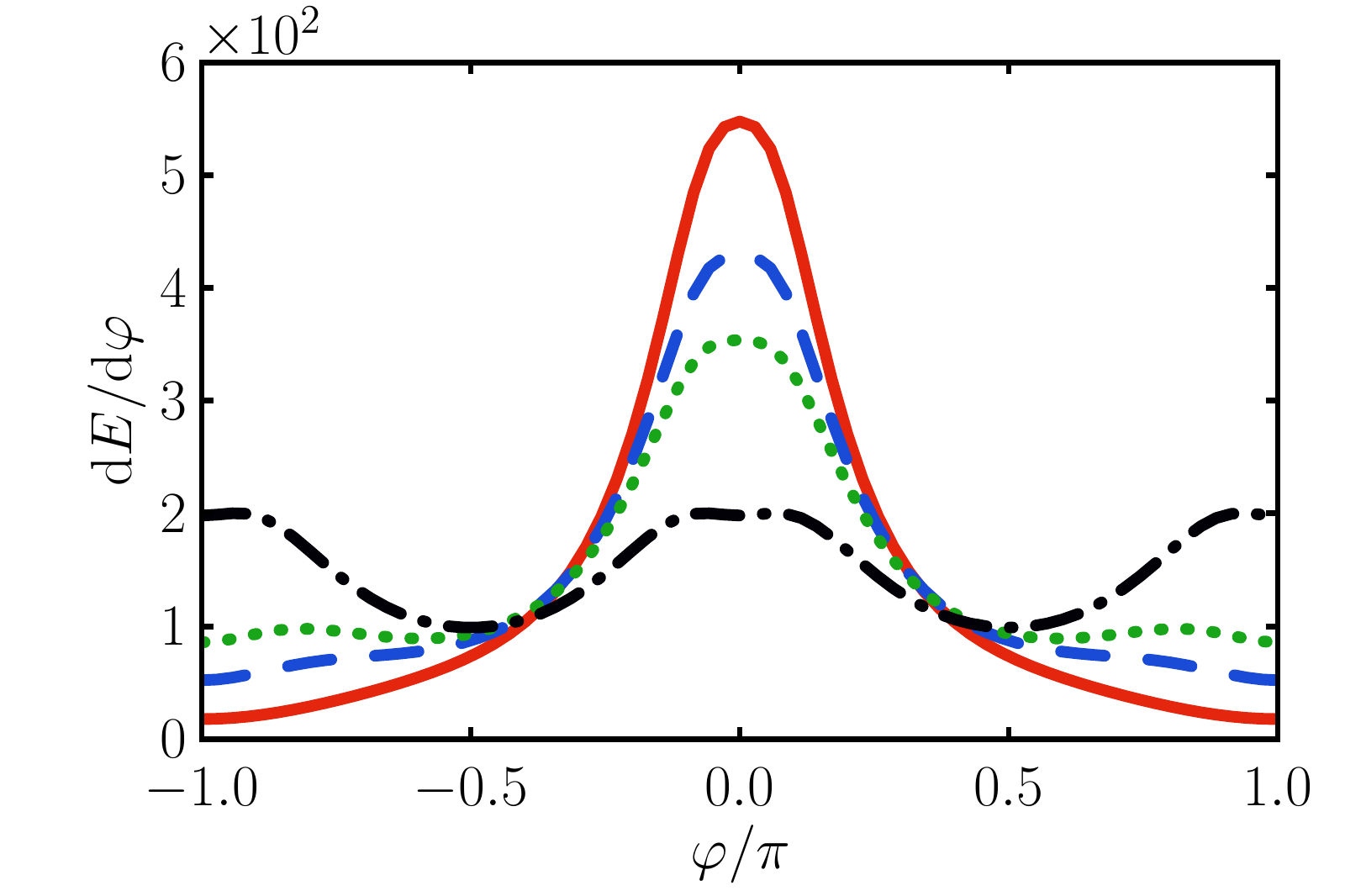}
\includegraphics[width=0.49\textwidth]{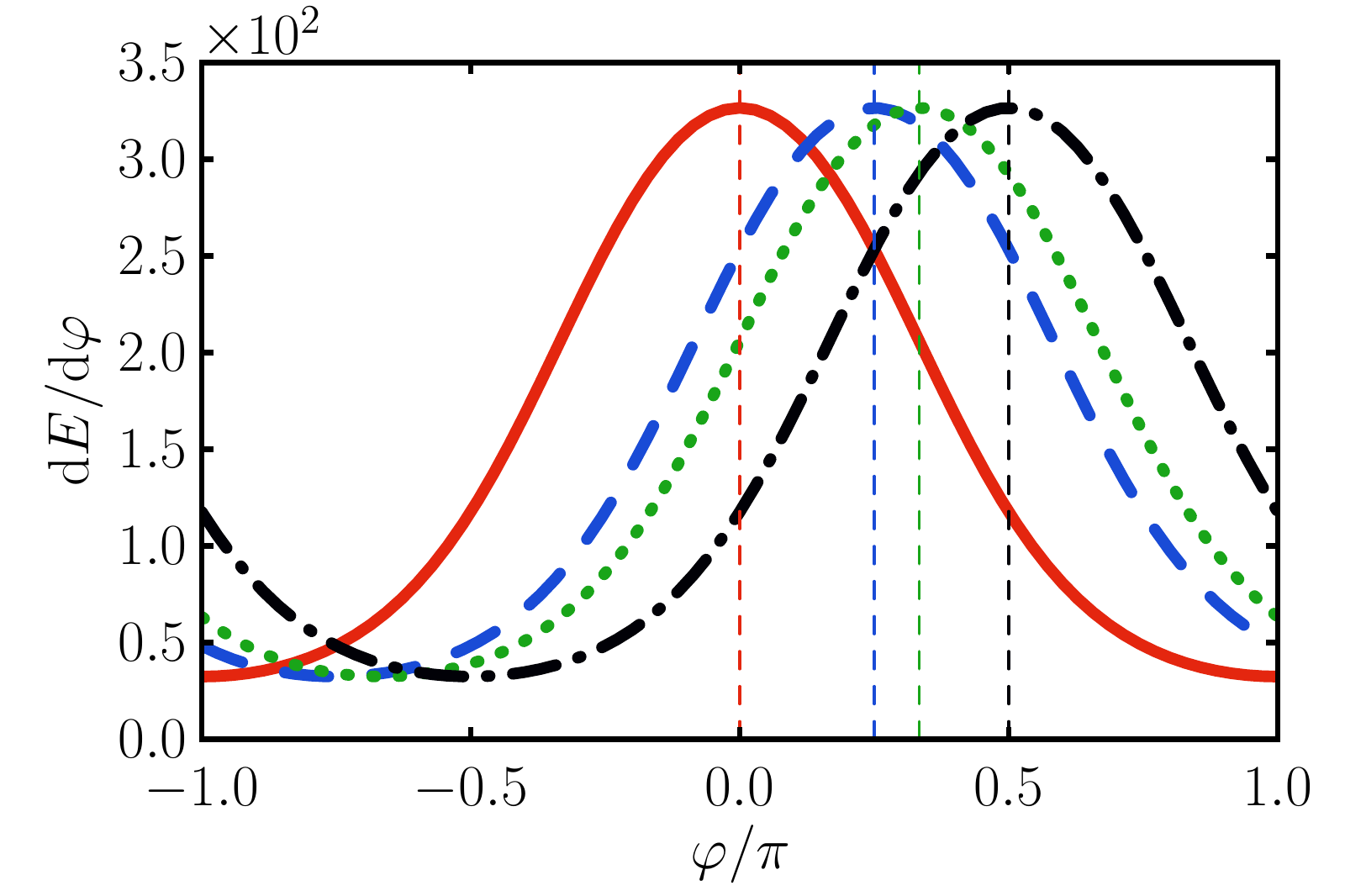}
\end{center}
\caption{Systematics of the dependence of azimuthal energy distribution
$\d E/\d \varphi$ on the CEP for 
	linear polarization in the $x$ direction ($\xi = 0$, left panel) and
	circular polarization ($\xi = \pi/4$, right panel).
The different curves in each panel are for various values of the CEP:
	$\phice = 0$ (red solid curve),
	$\pi/4$ (blue dashed),
	$\pi/3$ (green dotted) and
	$\pi/2$ (black dash-dotted).
	In all cases $a_0=2$ and $N=1$.
		}
\label{fig:phice}
\end{figure}
At low laser intensity, $a_0 \ll 1$, the shape of the azimuthal distributions does not depend on
the value of $\phice$.
However, for large values of $a_0$, there is a strong dependence
of the azimuthal energy distribution on the CEP.
For $\phice=0$ the emission probability has the strong unidirectional characteristics with a maximum at $\varphi=0$, since then the
laser vector potential is strongly antisymmetric and large values are achieved only in the direction
$\varphi=0$.
The shape of the spectrum changes upon increasing $\phice$ up to $\phice = \pi/2$ from the unidirectional emission to a bi-polar pattern for $\phice=\pi/2$ in the direction of the polarization three-vector $\boldsymbol \epsilon = \mathbf e_x$ of the laser, see left panel of Fig.~\ref{fig:phice}.
In the case of $\phice = \pi/2$, the laser vector potential is symmetric with respect to $\varphi=0$
and $\varphi=\pi$ in the sense that the maximum value of the vector potential in the left and right hemispheres has the same value. This symmetry is translated to the left-right symmetry of the azimuthal spectrum.
Upon increasing $\phice$ further up to $\phice=\pi$, one finds an unidirectional emission with the maximum at $\varphi=\pi$, i.e.~the opposite direction as for $\phice=\pi$, reflecting the
opposite sign of the vector potential in that case.
This sensitive dependence on the CEP $\phice$ for $a_0 \gtrsim 1$ is
the basis for an experimental access to the CEP proposed in
\cite{Mackenroth:PRL2010} for ultra-strong laser pulses, although it was discussed there in terms of
the polar angle spectra.

In the case of circular laser polarization, the shape of the spectrum does not depend on
the value of the CEP, but the position of the maximum of the distribution
does. This is exhibited in the right panel of Fig.~\ref{fig:phice}, where the
different curves represent the azimuthal spectra for various values of $\phice$.
The straight vertical lines mark the maxima of the respective distributions and are equal to
the value of the CEP.
Although the usual azimuthal symmetry, which is a typical characteristic of a long circularly polarized
laser pulse, is lost, a new symmetry arises.
In a short laser pulse, the Compton spectrum depends only on the difference $\varphi - \phice$, replacing the usual azimuthal symmetry.
This effectively reduces one degree of freedom of the parameter space, since the coordinate system
may always be oriented such that it corresponds to $\phice=0$ in our parametrization
of the laser pulse, i.~e.~the maximum of the vector potential occurs for $\varphi=0$.

This symmetry can be deduced analytically from the expressions \eqref{eq:probability.2} for the differential emission probability. In general, any of the four functions $\mathrsfs C_n$ entering 
$w$ depends on the CEP through 
their non-linear phase factors $f(\phi)$ (cf.~Eq.~\eqref{eq:f}).
Additional dependencies in the three functions $\mathrsfs C_\pm$ and $\mathrsfs C_2$
cancel from the expression \eqref{eq:probability.2} upon
specifying circular polarization, i.e.~for $\cos 2\xi = 0$.
For the non-linear phase $f(\phi)$ of the $\mathrsfs C_n$ we find for head-on collisions (cf.~Eq.~\eqref{eq:f})
that $\varphi$ and $\phice$ appear solely in the combination
\begin{align}
\epsilon_+\cdot p' \, e^{- i\phice} &= - \epsilon_+ \cdot k' \, e^{- i\phice} 
= \frac{\omega' \sin \vartheta}{\sqrt2} e^{\pm i (\varphi - \phice)}
\end{align}
proving that the differential probability $w$ depends only on the difference $\varphi - \phice$ and not 
on the individual values of $\varphi$ and $\phice$.

\subsubsection{Dependence on the laser polarization}

\begin{figure}[!t]
\begin{center}
\includegraphics[width=0.49\textwidth]{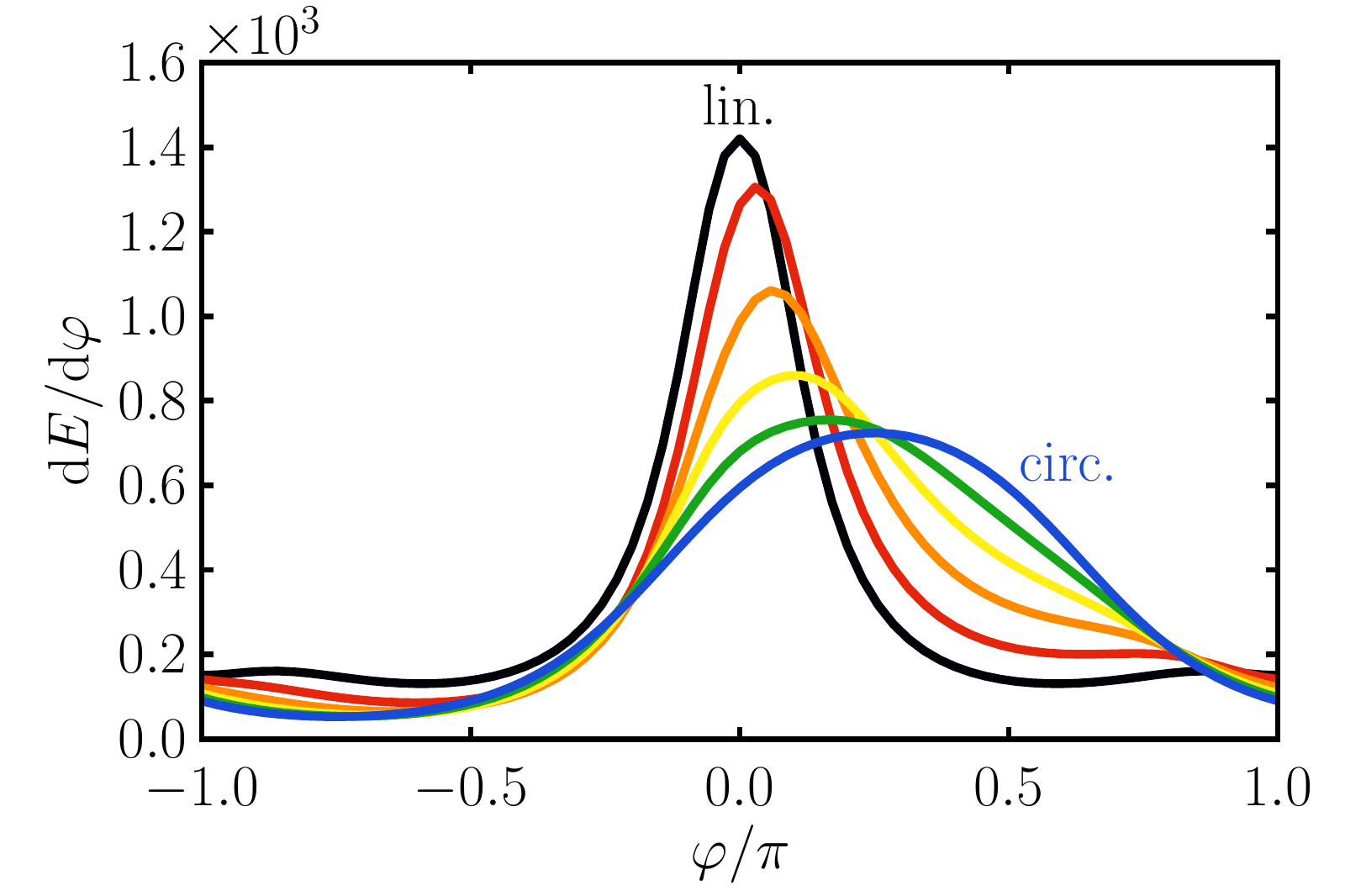}
\end{center}
\caption{Azimuthal emission spectra for $a_0=3$, $N=1$, $\phice = \pi/4$ and various
values of the laser polarization parameter starting with linear polarization
$\xi/\pi=0$ (black curve, ``lin.'')
and increasing the value of $\xi/\pi$ in steps of $0.05$ up to $\xi/\pi = 0.25$ corresponding to
circular polarization (blue curve, ``circ.'').
}
\label{fig:polarization}
\end{figure}
Hitherto we considered the extreme cases of linear ($\xi=0$) and circular ($\xi=\pi/4$)
laser polarization. Now, the impact of intermediate elliptic polarizations on the azimuthal distributions
is analysed.
The azimuthal emission spectra for a single-cycle laser pulse, $N=1$, are exhibited for $a_0=3$ and $\phice=\pi/4$ in Fig.~\ref{fig:polarization} for various laser
polarizations $\xi$ ranging from linear over elliptic to circular polarization.
Indeed, the azimuthal emission spectra show a characteristic dependence on the polarization of the laser pulse,
which gradually develops from the narrow unidirectional emission on the axis of polarization for linear polarization (black curve in Fig.~\ref{fig:polarization})
to the directional emission into the preferential direction $\varphi =  \phice = \pi/4$ for circular polarization (blue curve in Fig.~\ref{fig:polarization}).
The peak position of the distribution is shifted monotonically to the right with increasing value of $\xi$ and the shape of the distribution gradually changes.
For generically elliptic laser polarization, the azimuthal distributions are very asymmetric.

\subsection{Asymmetries of the azimuthal distributions}

\begin{figure}[!t]
\includegraphics[width=0.49\textwidth]{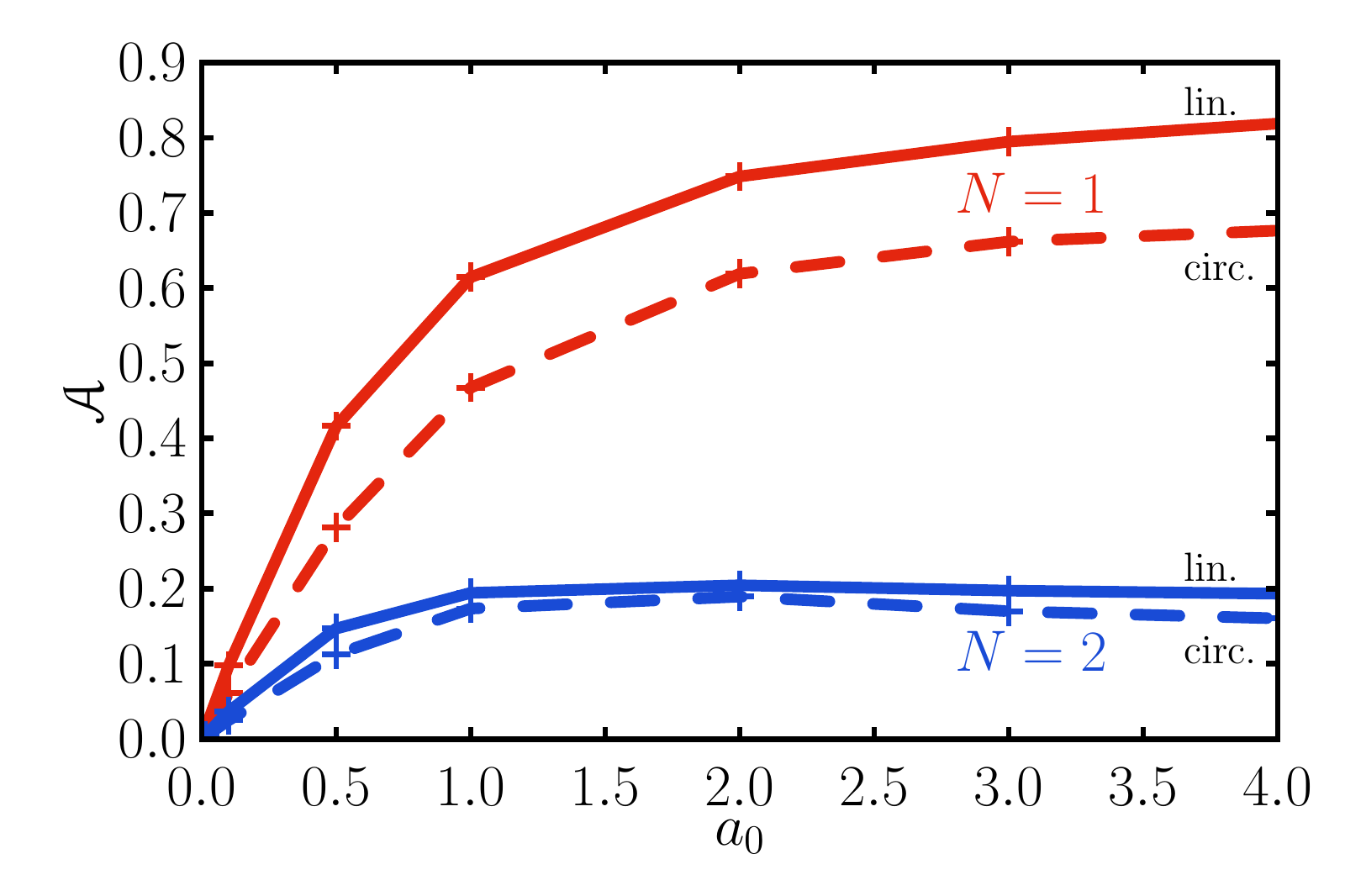}
\includegraphics[width=0.49\textwidth]{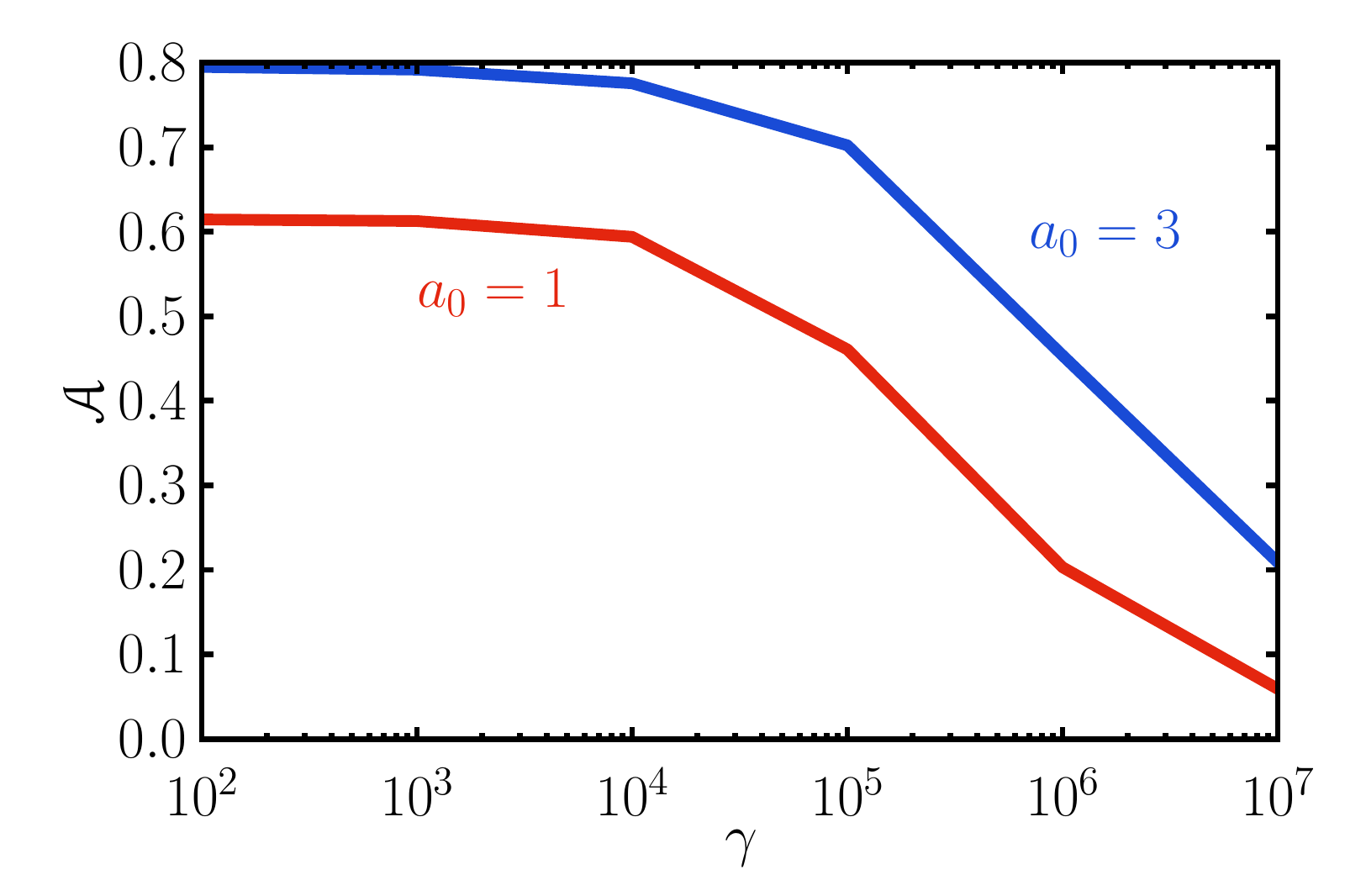}
\caption{Left panel: The azimuthal asymmetry $\mathscr A$ as
a function of the laser strength $a_0$ for a single-cycle pulse ($N=1$) and a two cycle pulse ($N=2$)
for $\gamma=100$. Solid (dashed) curves are for linear (circular) polarization. Right panel: Anisotropy $\mathscr A$ as a function of the initial electron energy $\gamma$ for $a_0=1$ (lower red curve) and $a_0=3$ (upper blue curve) and linear polarization.}
\label{fig:asymmetry.linear}
\end{figure}

The differential information in the azimuthal distribution $\d E/ \d \varphi$ may be
further condensed in the asymmetry $\mathscr A$ defined as
\begin{align}
\mathscr A &= \frac{ E_{\rightarrow} - E_{\leftarrow}}
{E_{\rightarrow} + E_{\leftarrow}} \,,
\end{align}
where
\begin{align}
E^{\rm lin}_{\rightarrow} &= \intop_{-\frac{\pi}{2}}^{\frac{\pi}{2}} \! \d \varphi \,  \frac{\d E}{\d \varphi} \,, & & E^{\rm lin}_{\leftarrow} = \intop_{\frac{\pi}{2}}^{\frac{3\pi}{2}} \! \d \varphi \,  \frac{\d E}{\d \varphi} \, 
\label{eq:def.asymmetry.linear}
\end{align}
denotes the energy emitted into the right ($\rightarrow$) and left ($\leftarrow$) hemispheres, respectively, for linear polarization.
For circular polarization, we exploit the symmetry in the variable $\varphi - \phice$
\begin{align}
E^{\rm circ}_{\rightarrow} &= \intop_{-\frac{\pi}{2} + \phice }^{\frac{\pi}{2} + \phice} \! \d \varphi \, \frac{\d E}{\d \varphi} \,, & &
E^{\rm circ}_{\leftarrow} = \intop_{\frac{\pi}{2} + \phice }^{\frac{3\pi}{2} + \phice} \! \d \varphi \, \frac{\d E}{\d \varphi} 
\label{eq:def.asymmetry.circular}
\end{align}
by shifting the orientation of the hemispheres by the value of the CEP.
Results for the asymmetry $\mathscr A$ as a function of $a_0$ are exhibited 
in Fig.~\ref{fig:asymmetry.linear} (left panel). We use $\phice=0$ which provides the largest
values of $\mathscr A$. From the results in Fig.~\ref{fig:asymmetry.linear}
it becomes clear that a large asymmetry is possible only for short pulses, e.g.~for $N=1$,
where linear polarization provides larger asymmetries than circular polarization.
Increasing the pulse length to $N=2$ leads to a reduction of the asymmetry, e.g.~for $a_0=4$
and linear polarization $\mathscr A$ drops from $0.82$ to $0.19$.

Within the classical theory of Thomson scattering based on electron trajectories,
which we briefly discuss in Appendix \ref{app:class}, the shape of the angular Compton spectrum depends only on the ratio $a_0/\gamma$ but not on the individual values $a_0$ and $\gamma$ \cite{Boca:PhysScr2011}. This scaling behaviour is violated within a quantum theoretical framework due to the electron recoil.
For increasing centre-of-mass energies, $\mathfrak s = (k+p)^2  \gg m^2$, the anisotropies
are gradually reduced until one finds an isotropic emission pattern
despite of the linear polarization of the laser field. In Compton backscattering of optical photons,
a higher centre-of-mass energy is related to a larger value of the electron Lorentz factor $\gamma$
by means of $\mathfrak s \simeq m^2 ( 1 + 4 \gamma^2 \omega/m)$.
In the right panel of Fig.~\ref{fig:asymmetry.linear}, the degradation of the asymmetry
with increasing values of $\gamma$ is exhibited.
For instance, for a high-energy electron beam with $\gamma=10^5$, which were used in the SLAC E-144 experiment \cite{Bamber:PRD1999},
the asymmetry has a value of $0.46$ for $a_0=1$ as compared to $0.61$ for the previously
discussed low-energy regime with $\gamma=100$ accessible, e.g.~at the HZDR.

A similar reduced dependence on the azimuthal angle for high centre-of-mass energies is also found in perturbative QED \cite{book:Jauch}. The corresponding expression for the emission cross section of a polarized photon is presented in Appendix \ref{app:pert}. As shown in
Eq.~\eqref{eq:pert.ultrarel}, the leading order in the ultra-relativistic limit $\mathfrak s \gg m^2$ is independent of the azimuthal angle $\varphi$.
The qualitative arguments that explain this behaviour are as follows:
For $\mathfrak s \gg m^2$, the centre-of-mass frame, in which the momenta of the outgoing electron and photon
are back-to-back,
moves rapidly with respect to both the laboratory frame and the frame where the electron is initially at rest.
Transforming back from the centre-of-mass frame the scattering angle of the
emitted photon is boosted close to the forward scattering direction to values
$\vartheta \simeq \pi - \gamma^{-1}$.
Close to forward scattering the coefficients of the non-linear phase integrals
are of the order $\alpha_+ = \mathcal O( a_0 / \gamma )$ and 
$\beta = \mathcal O (a_0^2/\gamma^2)$,
i.e.~both are small compared to unity. Thus, the emission of high frequency photons
with large values of the variable $s$, which are responsible for the asymmetries in the azimuthal emission patterns, are strongly suppressed.

\section{Total emitted Energy}

\label{sect:energy}
After having characterized the angular distribution of the energy integrated photon
intensity, we now briefly discuss the total amount of energy $E$ which is
radiated off in the non-linear Compton process.
For the situation considered here, characterized by $a_0 \ll \gamma$, we find
$E^{\rm cl} = 2 \gamma^2 \sigma_T \enflux$ within a classical framework for the
emission of radiation (i.e.~Thomson scattering, see Appendix \ref{app:class}).
The quantity $\sigma_T=\unit{665}{\milli\barn}$ denotes the Thomson cross section,
which is related to the classical electron radius $r_e$ via $\sigma_T=8\pi r_e^2/3$, and
$\enflux= \int \! \d x^+ \, n_i T^{i0}(\phi)$
is the primary energy flux, integrated over the light-front time $x^+$
with $n_i$ denoting the spatial components of the laser propagation direction.
The integrated primary energy flux $\enflux$ corresponds to the total laser
energy irradiated onto a unit area during a single pulse.
For the laser vector potential $A^\mu$ given in \eqref{eq:vector.potential}, we find the energy momentum tensor $T^{\mu\nu} = - k^\mu k^\nu  A' \cdot A'$. (The prime denotes the derivative w.r.t.~the laser phase $\phi$.)
Evaluating the derivative, the integrated energy flux can be written as
$\enflux = \frac{m^2  a_0^2 \omega }{2e^2 } \Delta \phi_{\rm eff}$
with the effective pulse length
$
\Delta \phi_{\rm eff} 
= \int_{-\infty}^\infty \! \d \phi \, \left[ (g^2 + g'^2) 
- \cos 2\xi \left\{ (g^2-g'^2) \cos 2(\phi + \phice) 
+ 2 g'g \sin 2(\phi+\phice) \right\} \right]$.
For longer pulses ($N>5$) it is suitable to approximate the integrand as $g^2(\phi)$,
such that we find $\Delta \phi_{\rm eff} \simeq 3\pi N/4$ for the specific pulse shape
\eqref{eq:envelope}.

\begin{figure}[!t]
\begin{center}
\includegraphics[width=0.49\textwidth]{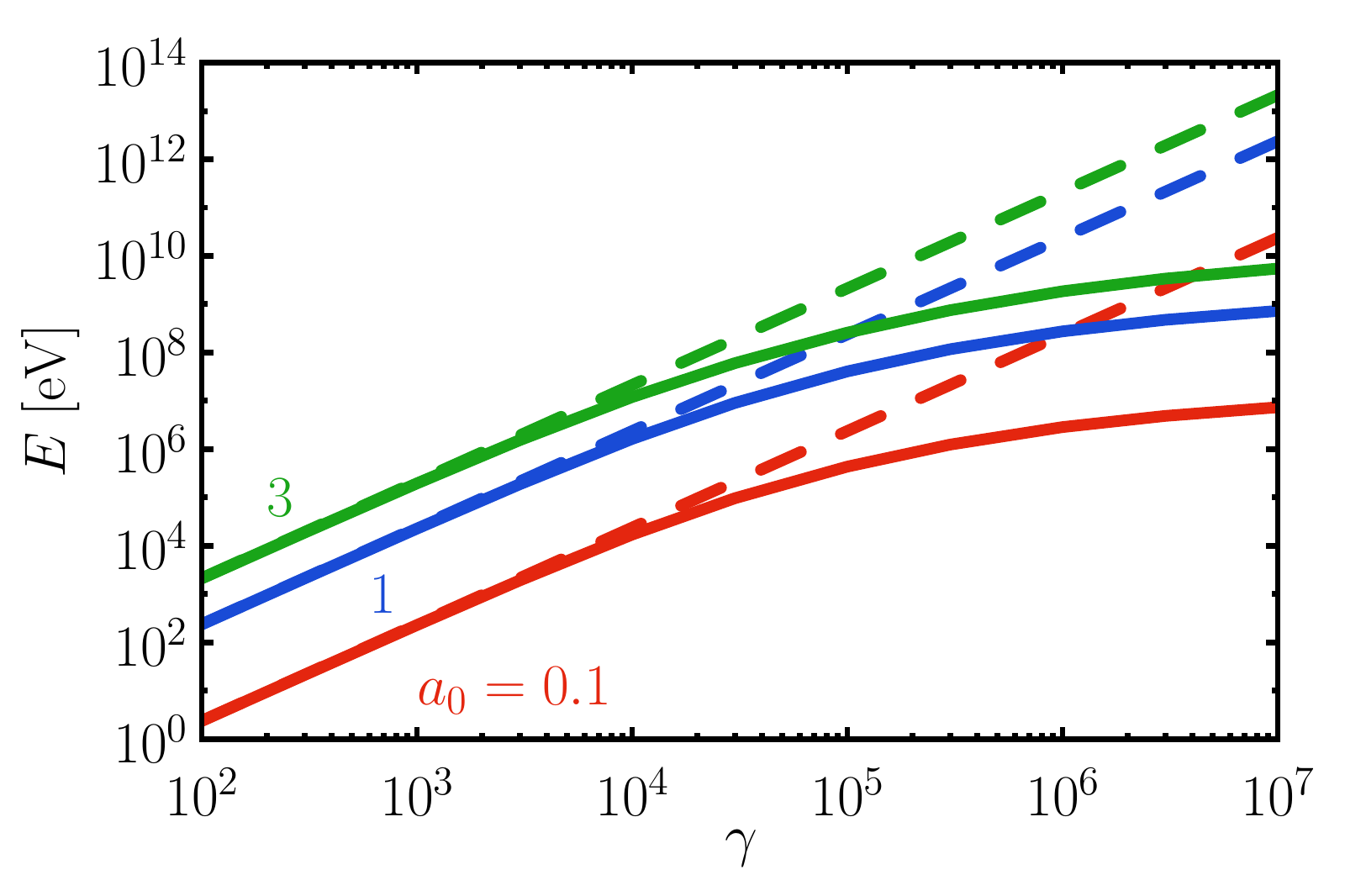}
\end{center}
\caption{Total emitted energy $E$ as a function of $\gamma$ for various values of $a_0=0.1$
(lower red curves), $1$ (middle blue curves) and $3$ (upper green curves). The solid curves depict
the full quantum result while the dashed curves refer to the classical approximation
$E^{\rm cl}$.
}
\label{fig:total}
\end{figure}

The emitted energy $E$ is exhibited in Fig.~\ref{fig:total} as a function of $\gamma$
for $a_0=0.1$, $1$ and $3$ from bottom to top.
For increasing values of $\gamma$, the total emitted energy stays below
the classical value $E^{\rm cl}$, which is a quantum effect due to the electron recoil.
The deviations start to become relevant for $\gamma$ between $10^3$ and $10^4$.
It is found that for higher laser intensity, the full quantum calculation deviates from the
classical expression already for lower values of $\gamma$. This is due to the emission
of high harmonics with large values of the variable $s$ in that case.
The relevant recoil parameter is $s \omega/m$, which may be large even if one is
still in the low-energy regime, characterized by $\omega/m \ll 1 $. 
This is in line with recently published results \cite{Dinu:2013} on the
total emitted energy for ultra-high laser intensity.

The total emitted energy is insensitive to the shape of the pulse.
The relevant quantity is the integrated flux of energy in the laser pulse $\enflux \propto a_0^2 N$.
This in contrast to the cross channel process of Breit-Wheeler type pair production,
where a strong enhancement of the emission probability
in ultra-short laser pulses was found near the threshold region
\cite{Titov:PRL2012,Nousch:PLB2012}.

\section{Discussion and Summary}
\label{sect.conclusion}

We study Compton backscattering of an ultra-short intense
laser pulse off an electron beam.
Non-linear and multi-photon effects are accounted for by working in the Furry picture
with laser dressed Volkov states.
Compton backscattering of a laser beam off an electron beam is interesting with respect
to the development of compact broadband pulsed X-ray radiation sources, as outlined in the Introduction.
A prerequisite of technical design studies is a comprehensive understanding of the elementary process.
To characterize the back-scattering spectrum we analyse angular distributions which
provide a clear signal of non-linear effects and show novel signatures in ultra-short pulses.
The angular distributions have the advantage not to be hampered by detector pile-ups as they represent an energy integrated distribution which is experimentally accessible in a calorimetric measurement.

It is shown that the energy integrated azimuthal spectra
show a strong dependence on both the laser polarization and the value of the
carrier envelope phase for ultra-short laser pulses.
The asymmetry for circular polarization is explained by the preferred direction where the vector potential reaches its
maximum value which breaks the usual azimuthal symmetry.
The azimuthal distribution displays a symmetry with respect to the difference of the azimuthal angle and the carrier envelope phase for ultra-short circularly polarized laser pulses.
This replaces the general azimuthal symmetry of infinitely long circularly polarized plane waves.

For non-phase-locked lasers one needs to perform single-shot measurements in
order see the asymmetries in the azimuthal spectra. 
For an electron bunch charge of $\unit{77}{\pico\coulomb}$ with gamma factor $\gamma=100$ interacting with a single-cycle laser pulse with intensity
${2\times 10^{18}}{\watt\per\centi\metre^2}$, we estimate the total number of photons as $2.8\times10^6$ with a total amount of deposited energy of $\unit{18}{\nano\joule}$ in the detector. This assumes an optimal overlap of the laser pulse and
the electron bunch.

The total emitted energy is proportional to the integrated primary energy flux in the laser pulse
in the low-energy regime and scales with the square of the primary electron energy within a
classical framework.
At high laser intensity we see deviations from the classical scaling behaviour
even in the Thomson regime due to the emission of
high harmonics, for which the quantum recoil is relevant.
The low-energy Thomson regime is representative for head-on collisions
of optical laser beams and mildly relativistic electron beams. For instance, the DRACO-ELBE constellation
at the HZDR fulfils this requirement.
Although the asymmetries for single-cycle pulses are presently out of reach of the
DRACO-ELBE system due too long laser pulses,
the transition from perpendicular emission for low laser intensity to
parallel emission with respect to the laser polarization for high intensity could well
be tested experimentally as a clear signal for non-linear effects in Compton scattering.

Our calculations apply for a section of transverse
laser beam profile where the curvature of the wave fronts is negligible since we rely on a plane
wave approximation. We restrict our analysis to mildly intense laser pulses with intensity
of the order of
$\unit{10^{19}}{\watt\per \centi\metre^2}$ which can be achieved in rather
large laser spots with sub-PW lasers, such that the effects of the curvature of
the laser fronts is negligible.
Multi-photon emission, e.g.~via the two-photon Compton process \cite{Seipt:PRD2012}, is expected to provide only a minor contribution to the radiation spectrum.

Focusing on the azimuthal distribution means considering the energy and
polar angle integrated spectra. In particular, the pile-ups prevent
in present-day detector technology the easy measurement of energy-differential spectra
for large incident photon fluxes,
even if they are interesting with respect to harmonics and their substructures.
The energy integrated quantities
$\d E/\d \Omega$ and $\d E/ \d \varphi$ are, in contrast, easily accessible experimentally
since they require an angularly resolved calorimetric measurement of the energy distributions.

We mention that all numerical results presented are for a particular temporal laser pulse shape.
It is straightforward to replace the employed function by other, possibly multi-parameter envelopes.

In summary we consider the non-linear Compton process of mildly relativistic electrons
interacting with medium-intense short and ultra-short laser pulses.
As a key to the radiation spectrum we propose the angular distributions
which are free of the notorious problem of pile-ups which may be faced in energy
differential spectra and provide clear signals of non-linear Compton scattering.
The impact of the laser intensity, the pulse length and the carrier envelope phase
on the energy-integrated azimuthal distributions is studied systematically.
Measurements of the azimuthal and polar angle distributions
may serve as useful tool to quantify both the electron beam and the laser beam parameters.

The parameter space which we have considered is representative
for many lasers worldwide in the sub-PW regime, i.e.~for intensity parameters $a_0 < 10$.
The ultra-intense regime deserves separate investigations. In contrast,
our consideration of ultra-short pulses with essentially one oscillation of the electromagnetic
field in a pulse is at the limit of present-day technological feasibility. The peculiarities, such as the strong asymmetry of the azimuthal distribution for circularly polarized laser radiation or the strong one-side
asymmetry for linearly polarized laser radiation, disappears quickly for more
oscillations of the electromagnetic field.
Therefore, such measures are particularly suitable tools for the characterization ultra-short pulses.

\begin{acknowledgements}
The authors thank A.~I.~Titov, T.~Heinzl, A.~Otto, T.~Nousch and T.~E.~Cowan for continuous useful
discussions and collaboration.
The support by R.~Sauerbrey, T.~St\"ohlker and S.~Fritzsche is gratefully acknowledged.
G.~Paulus pointed out the important role of the carrier envelope phase which we subsequently
considered here.
\end{acknowledgements}

\appendix

\section{Classical Radiation: Thomson Scattering}
\label{app:class}

In the classical picture, which is valid if the electron recoil during the interaction
is negligible \cite{Seipt:PRA2011}, an ultra-relativistic electron, moving in an external electromagnetic wave field, emits radiation only into directions which are swept by the instantaneous tangent vector to its trajectory, i.e.~the
spatial part $\mathbf u$ of the four-velocity $u^\mu(\tau) = \d  x^\mu / \d \tau = (\gamma,\mathbf u)$, with proper time $\tau$.
For plane wave laser fields, the orbit reads in terms of the laser vector potential $A^\mu$
\begin{align}
u^\mu(\tau) &= u_0^\mu - \frac{e}{m} A^\mu  + k^\mu 
\left( \frac{e}{m}\frac{ u_0 \cdot A}{k\cdot u_0}  - \frac{e^2}{m^2}\frac{ A \cdot A}{2k\cdot u_0} \right)
\label{eq:orbit} \,.
\end{align}
The differential radiated power
of an electron moving
on an arbitrary trajectory reads (cf.~\cite{book:Jackson}, Eq.~(14.37))
\begin{align}
\left.
\frac{\d E (t')}{\d \Omega \d t'  } = \frac{\alpha}{4\pi}
\frac{
	\left| 
	\mathbf n' \times \left[
	 (\mathbf n' - \mathbf v) \times \frac{\d \mathbf v}{\d t}
	\right]
	\right|^2}{(1 - \mathbf n' \cdot \mathbf v)^6} \right|_{\rm ret} \,.
\end{align}
Integration over observation time $t'$ and a change of the integration variables retarded time $t$ yields
\begin{align}
\frac{\d E }{\d \Omega} = \frac{\alpha}{4\pi}
\int \! \d t \,
\frac{
	\left| 
	\mathbf n' \times \left[
	 (\mathbf n' - \mathbf v) \times \frac{\d \mathbf v}{\d t}
	\right]
	\right|^2}{(1 - \mathbf n' \cdot \mathbf v)^5} \,.
\end{align}
Now, we replace the velocity $\mathbf v = \d \mathbf x/\d t$ by the four-velocity $u^\mu$
and change the integration variable from $t$ to the laser phase $\phi$
via $\d t = \frac{\d t}{\d \tau} \frac{\d \tau}{\d \phi} \d \phi = \frac{\gamma}{k\cdot u_0} \d \phi$.
We have $\mathbf v = \mathbf u / \gamma$ and
${\d \mathbf v}/{\d t} 
= ({\dot{\mathbf u} \gamma - \mathbf u \dot \gamma})/{\gamma^3}$,
where the dot denotes the derivative w.r.t.~proper time $\tau$.
Using these relations we obtain
\begin{align}
\frac{\d E }{\d \Omega} = \frac{\alpha}{4\pi k\cdot u_0}
\int \! \d \phi \,
\frac{
	\left| 
	\mathbf n' \times \left[
	 (\mathbf n' \gamma - \mathbf u) \times ( \dot{\mathbf u} \gamma - \mathbf u \dot{\gamma} )
	\right]
	\right|^2}{\gamma^2 (n' \cdot u)^5} 
	\,.
\end{align}
The evaluation of the square of the double cross product yields,
exploiting the relativistic constraints $u\cdot u = 1$ and $\dot u \cdot u = 0$,
the result
\begin{align}
	\frac{1}{\gamma^2}\left| 
	\mathbf n' \times \left[
	 (\mathbf n' \gamma - \mathbf u) \times ( \dot{\mathbf u} \gamma - \mathbf u \dot{\gamma} )
	\right]
	\right|^2 & =
	- (n' \cdot \dot u)^2  - (n' \cdot u)^2 \dot u^2 \nonumber \\
		& =
- \left( \frac{\d }{\d \tau}  n'\cdot u \, u^\mu  \right)^2	\,.
	\label{eq:double.cross.square}
\end{align}
Thus, the radiated energy is given by the expression
\begin{align}
\frac{\d E }{\d \Omega} 
	=
	- \frac{\alpha  k\cdot u_0 }{4\pi}
\int \!  
\frac{\d \phi}{ (n' \cdot u)^5} 
	\left(\frac{\d}{\d \phi}  n' \cdot  u \, u^\mu \right)^2 
	\,.
	\label{eq:E.class}
\end{align}
Using the trajectory as a function of the laser vector potential in Eq.~\eqref{eq:orbit}
allows to calculate
the angular radiation spectrum. Our result \eqref{eq:E.class} compares well to the result in \cite{Boca:PhysScr2011}. However, we provide a covariant expression in terms
of the orbit $u^\mu$ in a plane wave \eqref{eq:orbit} and
furthermore, we do not rely on the approximation $\mathbf u \parallel \mathbf n'$ to simplify the
expression for the emitted energy distribution.

To calculate the total amount of radiated energy we may either integrate
\eqref{eq:E.class} over the full solid angle
or alternatively use the Larmor formula for the 
emitted power (cf.~\cite{book:Jackson}, Eq.~(14.24))
\begin{align}
\frac{\d E^{\rm cl}}{\d t} = - \frac{1}{6\pi} \frac{e^2}{m^2} \, \dot p \cdot \dot p \,.
\label{eq:larmor}
\end{align}
With the classical equations of motion for the electron $\dot p^\mu = \frac{e}{m} F^{\mu\nu} p_\nu$,
where $F_{\mu\nu} = \partial_\mu A_\nu - \partial_\nu A_\mu$ is the electromagnetic field strength tensor of the laser field, the emitted power can be expressed as
\begin{align}
\frac{\d E^{\rm cl}}{\d t} =  \frac{1}{6\pi} \frac{e^4}{m^4} p_\mu F^{\mu\nu} F_{\nu}^{\ \kappa} p_\kappa
= \frac{1}{6\pi} \frac{e^4}{m^4} p_\mu T^{\mu\nu} p_\nu
\,, 
\end{align}
where the last step is valid for a null background field, as is used here. Evaluating the energy momentum tensor for the vector potential \eqref{eq:vector.potential} one finds
$T^{\mu\nu} = k^\mu k^\nu   T^{00} / \omega^2$. 
Thus, the emitted power is proportional to the energy density of the background field
\begin{align}
\frac{\d E^{\rm cl}}{\d t} = \frac{1}{6\pi} \frac{e^4  (k\cdot p)^2}{m^4 \omega^2}T^{00} 
\,.\label{eq:power}
\end{align}
The total energy emitted in such a process is the coordinate time integral of \eqref{eq:power}, which needs to be transformed into
an integral over the laser phase as above, yielding
\begin{align}
 E^{\rm cl} 
=   \frac{8}{3} \pi r_e^2   \frac{k\cdot u_0}{\omega^2}\int \d \phi \,\gamma(\phi) T^{00}(\phi)
 \,. \label{eq:classical.energy}
\end{align}
For head-on collisions with $\gamma_0 \gg1$, one finds
$\gamma(\phi) \simeq \gamma_0 [ 1 + \mathcal O ( a_0/\gamma_0)^2  ]$ indicating that for moderately intense lasers with $a_0 \ll \gamma_0 $,
the initial value $\gamma^0$ is the leading-order
contribution which can be taken out of the integral. Thus, the emitted energy
is proportional to the integrated energy flux of the laser pulse $\enflux$,
and
$E^{\rm cl} \simeq 2 \gamma_0^2 \sigma_T  \enflux$.
The next-to-leading order gives a positive correction, whereas the quantum corrections,
that scale with the quantum non-linearity parameter $\chi$,
have a negative sign \cite{Ritus:JSLR1985}.

\section{Perturbative Compton Scattering with Linearly Polarized Photons}
\label{app:pert}

In perturbative QED, the cross section for Compton scattering of unpolarized electrons
and polarized photons reads in the rest frame of the incoming electron \cite{book:Jauch}
\begin{align}
\frac{\d \sigma}{\d \Omega} &= \frac{r_e^2}{4} 
\left( \frac{\omega'}{\omega}\right)^2
\left( 
\frac{\omega'}{\omega} + \frac{\omega}{\omega'} -2 + 4 (\boldsymbol \epsilon \cdot \boldsymbol \epsilon')^2
\right) \,,
\end{align}
where $\boldsymbol \epsilon$ ($\boldsymbol \epsilon'$) denotes the polarization vector of the
incoming (outgoing) photon and
\begin{align}
\omega' &= \frac{\omega}{1+ \nu (1+\cos \vartheta)}
\end{align}
is the frequency of the outgoing photon as a function of the scattering angle $\vartheta$ (note that in our convention forward scattering is denoted by $\vartheta=\pi$);
$r_e = \alpha/m$ is the classical electron radius and $\nu= \omega/m$
denotes normalized laser frequency
such that the centre-of-mass energy squared is $\mathfrak s = m^2(1+2\nu)$.
Specifying $\boldsymbol \epsilon = \mathbf e_x$ and summing over the final photon polarizations
we obtain
\begin{align}
\frac{\d \sigma}{\d \Omega} &= \frac{r_e^2}{2} 
\left( \frac{\omega'}{\omega}\right)^2
\left( 
\frac{\omega'}{\omega} + \frac{\omega}{\omega'} - 2 \cos^2 \varphi \sin^2 \vartheta \right) \,.
\end{align}
The integration over the polar angle $\vartheta$ yields the azimuthal distribution
\begin{align}
\frac{\d\sigma}{\d\varphi} &=
\frac{r_e^2}{2} \left( 
\frac{\log (1+2\nu)}{\nu} + 2\frac{(1+\nu)}{(1+2\nu)^2}
-\frac{4 \cos^2 \varphi}{\nu^3} [ (1+\nu ) \log(1+2\nu) -2\nu]
\right)\,. \label{eq:pert.cs.phi}
\end{align}
The non-relativistic ($\nu \ll1$) and ultra-relativistic ($\nu\gg1$) limits read
\begin{align}
\frac{\d\sigma}{\d\varphi} &=
\frac{r_e^2}{2} 
\left[
\left(4 - \frac{8}{3} \cos^2\varphi \right)
-2\nu \left(4 - \frac{8}{3} \cos^2\varphi \right) 
\right] + \mathcal O(\nu^2), & & \nu \ll1 \\
\frac{\d\sigma}{\d\varphi} &=
\frac{r_e^2}{2} 
\left[
\frac{1}{\nu}\left(\frac12 + \log 2\nu \right)
+ \frac{1}{\nu^2} \left(\frac12 +8 \cos^2\varphi  - 4 \cos^2 \varphi \log 2\nu \right) 
\right] + \mathcal O(\nu^{-3}), & & \nu \gg1 \,.
\label{eq:pert.ultrarel}
\end{align}

The leading-order term of the non-relativistic limit of the perturbative cross section gives the same azimuthal distribution as in the general expression $\d W/\d \varphi$
(or $\d E/\d \varphi$) for weak short laser pulses, i.e.~in the limit $a_0 \ll 1$.
On the contrary, the leading order of the ultra-relativistic limit turns out to be independent of the
azimuthal angle $\varphi$; an azimuthal dependence enters first at next-to-leading order.

\end{document}